\documentclass[prd,showkeys,floatfix,twocolumn,amsmath,amssymb]{revtex4-2}
\usepackage{graphicx}
\usepackage{epstopdf}
\usepackage{multirow}
\usepackage{subfigure}
\usepackage[colorlinks,citecolor=blue,anchorcolor=red,menucolor=red,linkcolor=red,filecolor=red,runcolor=red,urlcolor=blue,frenchlinks=red]{hyperref}
\newcommand{\feynp}[1]{#1\kern-0.45em/}

\allowdisplaybreaks[3]

\begin{document}
\title{Unified spectroscopy of $P$-wave flavor-sextet heavy baryons from QCD sum rules}
%

\author{Shu-Wei Zhang$^1$}
\author{Xuan Luo$^1$}
\author{Hua-Xing Chen$^1$}
\email{hxchen@seu.edu.cn}
\author{Hui-Min Yang$^2$}
\email{hmyang@pku.edu.cn}

\affiliation{
$^1$School of Physics, Southeast University, Nanjing 210094, China
\\
$^2$School of Physics, Henan Normal University, Henan 453007, China
}

\begin{abstract}
We investigate the $P$-wave flavor-sextet charmed and bottom baryons within heavy quark effective theory. A key motivation is provided by the recent evidence for the $\Xi_c(2882)^0$, which completes a sequence of four narrow $\Xi_c$ structures together with the $\Xi_c(2923)^0$, $\Xi_c(2939)^0$, and $\Xi_c(2965)^0$. Their characteristic mass-splitting pattern closely parallels that of the $\Omega_c(3000)^0$, $\Omega_c(3050)^0$, $\Omega_c(3066)^0$, and $\Omega_c(3090)^0$ states, providing strong constraints on their spectroscopic assignments. We classify the seven $P$-wave states in each of the $\Sigma_Q$, $\Xi_Q^\prime$, and $\Omega_Q$ sectors ($Q=c,b$), calculate their masses and intra-doublet mass splittings using QCD sum rules, and study their strong decays using light-cone sum rules. Configuration mixing between states with the same quantum numbers is also investigated. The combined analysis favors a common interpretation of the four narrow $\Xi_c$ and the four lowest narrow $\Omega_c$ structures in terms of the corresponding $\lambda$-mode excitations. Extending the same framework to the bottom sector, we obtain a coherent picture of the observed $\Sigma_b$, $\Xi_b$, and $\Omega_b$ structures. In particular, the $\Sigma_b(6097)$, $\Xi_b(6227)$, and $\Omega_b(6350)$ structures may each contain an unresolved pair of nearby $P$-wave states. We also predict two additional $\Sigma_b$ states, two additional $\Xi_b^\prime$ states, and one additional $\Omega_b$ state, all of which are expected to be relatively narrow and remain to be identified experimentally.
\end{abstract}



\maketitle

\pagenumbering{arabic}

\section{Introduction}
\label{sec:introduction}

Singly heavy baryons provide an ideal laboratory for studying the nonperturbative dynamics of QCD and the fine structure of hadron spectra~\cite{pdg,Chen:2016spr,Chen:2022asf,Luo:2025sns,Copley:1979wj,Karliner:2008sv}. In the heavy quark limit, the heavy quark behaves approximately as a static color source, while the light quarks and gluons constitute the dynamical light degrees of freedom around it. In this sense, a singly heavy baryon resembles the QCD analogue of a hydrogen atom~\cite{Korner:1994nh,Manohar:2000dt,Bianco:2003vb,Klempt:2009pi}. Heavy quark symmetry therefore provides a natural organizing principle for the excitation spectrum and, in particular, relates the charm and bottom sectors. Motivated by this perspective, we recently investigated the quantum numbers of the excited $\Xi_c^\prime$ and $\Omega_c$ baryons together with the $P$-wave $\Sigma_c$ spectrum~\cite{Luo:2026elv}. The present study incorporates and substantially extends that analysis to provide a unified description of the $P$-wave flavor-sextet charmed and bottom baryons.

Substantial experimental progress has been made in the spectroscopy of singly charmed baryons, providing a rich set of structures relevant to the $P$-wave flavor-sextet spectrum:
\begin{itemize}

\item The $\Sigma_c(2800)$ structures were first observed by the Belle Collaboration~\cite{Belle:2004zjl}. Their measured masses and widths are
\begin{align}
M_{\Sigma_c(2800)^{++}}
&=2801^{+4}_{-6}~{\rm MeV},
\nonumber\\
\Gamma_{\Sigma_c(2800)^{++}}
&=75^{+18+12}_{-13-11}~{\rm MeV},
\nonumber\\
M_{\Sigma_c(2800)^{+}}
&=2792^{+14}_{-5}~{\rm MeV},
\nonumber\\
\Gamma_{\Sigma_c(2800)^{+}}
&=62^{+37+52}_{-23-38}~{\rm MeV},
\nonumber\\
M_{\Sigma_c(2800)^{0}}
&=2806^{+5}_{-7}~{\rm MeV},
\nonumber\\
\Gamma_{\Sigma_c(2800)^{0}}
&=72^{+22}_{-15}~{\rm MeV}.
\label{eq:sigmac2800-exp}
\end{align}
More recently, the LHCb Collaboration performed an amplitude analysis of the $B^-\to\Lambda_c^+\bar p\pi^-$ decay and observed a new $\Sigma_c(2900)^0$ structure~\cite{LHCb:2026nzu}. Two statistically indistinguishable families of amplitude solutions were obtained, yielding
\begin{align}
M_{\Sigma_c(2900)^0}
&=2908\pm5\pm9~{\rm MeV},
\qquad {\rm [Group~A]},
\nonumber\\
\Gamma_{\Sigma_c(2900)^0}
&=175\pm8\pm23~{\rm MeV},
\qquad {\rm [Group~A]},
\nonumber\\
M_{\Sigma_c(2900)^0}
&=2914\pm3\pm8~{\rm MeV},
\qquad {\rm [Group~B]},
\nonumber\\
\Gamma_{\Sigma_c(2900)^0}
&=92\pm6\pm23~{\rm MeV},
\qquad {\rm [Group~B]}.
\label{eq:sigmac2900-exp}
\end{align}
The observation of the $\Sigma_c(2900)^0$ further enriches the experimental spectrum in the mass region relevant to the $P$-wave $\Sigma_c$ excitations.

\item Earlier experimental studies reported the $\Xi_c(2930)$ and $\Xi_c(2970)$ structures in several decay processes~\cite{Belle:2006edu,BaBar:2007xtc,BaBar:2007zjt,Belle:2020tom}. Representative measurements of their masses and widths are
\begin{align}
M_{\Xi_c(2930)^+}
&=2942.3\pm4.4\pm1.5~{\rm MeV},
\nonumber\\
\Gamma_{\Xi_c(2930)^+}
&=14.8\pm8.8\pm2.5~{\rm MeV},
\nonumber\\
M_{\Xi_c(2930)^0}
&=2929.7^{+2.8}_{-5.0}~{\rm MeV},
\nonumber\\
\Gamma_{\Xi_c(2930)^0}
&=26\pm8~{\rm MeV},
\nonumber\\
M_{\Xi_c(2970)^+}
&=2966.34^{+0.17}_{-1.00}~{\rm MeV},
\nonumber\\
\Gamma_{\Xi_c(2970)^+}
&=20.9^{+2.4}_{-3.5}~{\rm MeV},
\nonumber\\
M_{\Xi_c(2970)^0}
&=2970.9^{+0.4}_{-0.6}~{\rm MeV},
\nonumber\\
\Gamma_{\Xi_c(2970)^0}
&=28.1^{+3.4}_{-4.0}~{\rm MeV}.
\label{eq:xic-old-exp}
\end{align}
A major development came in 2020, when the LHCb Collaboration resolved three narrow $\Xi_c^0$ structures, $\Xi_c(2923)^0$, $\Xi_c(2939)^0$, and $\Xi_c(2965)^0$, in the $\Lambda_c^+K^-$ invariant-mass spectrum~\cite{LHCb:2020iby}. Their masses and widths were measured to be
\begin{align}
M_{\Xi_c(2923)^0}
&=2923.04\pm0.25\pm0.20\pm0.14~{\rm MeV},
\nonumber\\
\Gamma_{\Xi_c(2923)^0}
&=7.1\pm0.8\pm1.8~{\rm MeV},
\nonumber\\
M_{\Xi_c(2939)^0}
&=2938.55\pm0.21\pm0.17\pm0.14~{\rm MeV},
\nonumber\\
\Gamma_{\Xi_c(2939)^0}
&=10.2\pm0.8\pm1.1~{\rm MeV},
\nonumber\\
M_{\Xi_c(2965)^0}
&=2964.88\pm0.26\pm0.14\pm0.14~{\rm MeV},
\nonumber\\
\Gamma_{\Xi_c(2965)^0}
&=14.1\pm0.9\pm1.3~{\rm MeV}.
\label{eq:xic-exp}
\end{align}
Subsequently, an LHCb study of the $B^-\to\Lambda_c^+\bar\Lambda_c^-K^-$ decay found evidence for another $\Xi_c^0$ structure~\cite{LHCb:2022vns}, denoted as $\Xi_c(2882)^0$ in the present study, whose mass and width were measured to be
\begin{align}
M_{\Xi_c(2882)^0}
&=2881.8\pm3.1\pm8.5~{\rm MeV},
\nonumber\\
\Gamma_{\Xi_c(2882)^0}
&=12.4\pm5.3\pm5.8~{\rm MeV}.
\label{eq:xic2882-exp}
\end{align}

\item In 2017, the LHCb Collaboration observed five narrow excited $\Omega_c^0$ structures, $\Omega_c(3000)^0$, $\Omega_c(3050)^0$, $\Omega_c(3066)^0$, $\Omega_c(3090)^0$, and $\Omega_c(3119)^0$, in the $\Xi_c^+K^-$ invariant-mass spectrum~\cite{LHCb:2017uwr}:
\begin{align}
M_{\Omega_c(3000)^0}
&=3000.4\pm0.2\pm0.1^{+0.3}_{-0.5}~{\rm MeV},
\nonumber\\
\Gamma_{\Omega_c(3000)^0}
&=4.5\pm0.6\pm0.3~{\rm MeV},
\nonumber\\
M_{\Omega_c(3050)^0}
&=3050.2\pm0.1\pm0.1^{+0.3}_{-0.5}~{\rm MeV},
\nonumber\\
\Gamma_{\Omega_c(3050)^0}
&=0.8\pm0.2\pm0.1~{\rm MeV},
\nonumber\\
M_{\Omega_c(3066)^0}
&=3065.6\pm0.1\pm0.3^{+0.3}_{-0.5}~{\rm MeV},
\nonumber\\
\Gamma_{\Omega_c(3066)^0}
&=3.5\pm0.4\pm0.2~{\rm MeV},
\nonumber\\
M_{\Omega_c(3090)^0}
&=3090.2\pm0.3\pm0.5^{+0.3}_{-0.5}~{\rm MeV},
\nonumber\\
\Gamma_{\Omega_c(3090)^0}
&=8.7\pm1.0\pm0.8~{\rm MeV},
\nonumber\\
M_{\Omega_c(3119)^0}
&=3119.1\pm0.3\pm0.9^{+0.3}_{-0.5}~{\rm MeV},
\nonumber\\
\Gamma_{\Omega_c(3119)^0}
&=1.1\pm0.8\pm0.4~{\rm MeV}.
\label{eq:omegac-exp}
\end{align}
Several of these structures were subsequently confirmed by the Belle Collaboration~\cite{Belle:2017ext} and further investigated by LHCb in the $\Omega_b^-\to\Xi_c^+K^-\pi^-$ decay~\cite{LHCb:2021ptx}.

\end{itemize}

The bottom sector has also witnessed substantial experimental progress, with several structures observed in the mass region relevant to the $P$-wave flavor-sextet spectrum:
\begin{itemize}

\item In 2018, the LHCb Collaboration observed the $\Sigma_b(6097)^\pm$ structures in the $\Lambda_b^0\pi^\pm$ invariant-mass spectra~\cite{LHCb:2018haf}. Their masses and widths were measured to be
\begin{align}
M_{\Sigma_b(6097)^+}
&=6095.8\pm1.7\pm0.4~{\rm MeV},
\nonumber\\
\Gamma_{\Sigma_b(6097)^+}
&=31.0\pm5.5\pm0.7~{\rm MeV},
\nonumber\\
M_{\Sigma_b(6097)^-}
&=6098.0\pm1.7\pm0.5~{\rm MeV},
\nonumber\\
\Gamma_{\Sigma_b(6097)^-}
&=28.9\pm4.2\pm0.9~{\rm MeV}.
\label{eq:sigmab-exp}
\end{align}

\item In the same year, the LHCb Collaboration observed the $\Xi_b(6227)^-$ structure in both the $\Lambda_b^0K^-$ and $\Xi_b^0\pi^-$ invariant-mass spectra~\cite{LHCb:2018vuc}. Its mass and width were measured to be
\begin{align}
M_{\Xi_b(6227)^-}
&=6226.9\pm2.0\pm0.3\pm0.2~{\rm MeV},
\nonumber\\
\Gamma_{\Xi_b(6227)^-}
&=18.1\pm5.4\pm1.8~{\rm MeV}.
\label{eq:xib-exp}
\end{align}

\item In 2020, the LHCb Collaboration observed four narrow $\Omega_b^-$ structures, $\Omega_b(6316)^-$, $\Omega_b(6330)^-$, $\Omega_b(6340)^-$, and $\Omega_b(6350)^-$, in the $\Xi_b^0K^-$ invariant-mass spectrum~\cite{LHCb:2020tqd}. Their masses and widths were measured to be
\begin{align}
M_{\Omega_b(6316)^-}
&=6315.64\pm0.31\pm0.07\pm0.50~{\rm MeV},
\nonumber\\
\Gamma_{\Omega_b(6316)^-}
&<2.8~{\rm MeV},
\nonumber\\
M_{\Omega_b(6330)^-}
&=6330.30\pm0.28\pm0.07\pm0.50~{\rm MeV},
\nonumber\\
\Gamma_{\Omega_b(6330)^-}
&<3.1~{\rm MeV},
\nonumber\\
M_{\Omega_b(6340)^-}
&=6339.71\pm0.26\pm0.05\pm0.50~{\rm MeV},
\nonumber\\
\Gamma_{\Omega_b(6340)^-}
&<1.5~{\rm MeV},
\nonumber\\
M_{\Omega_b(6350)^-}
&=6349.88\pm0.35\pm0.05\pm0.50~{\rm MeV},
\nonumber\\
\Gamma_{\Omega_b(6350)^-}
&=1.4^{+1.0}_{-0.8}\pm0.1~{\rm MeV}.
\label{eq:omegab-exp}
\end{align}

\end{itemize}
Taken together, these experimental observations provide a timely opportunity to explore whether the charmed and bottom spectra can be described within a unified spectroscopic framework based on heavy-quark symmetry.

The rapidly expanding experimental landscape has stimulated extensive theoretical studies of excited singly heavy baryons. Their spectroscopic properties have been investigated within constituent and relativistic quark models~\cite{Capstick:1985xss,Chen:2007xf,Garcilazo:2007eh,Ebert:2007nw,Roberts:2007ni,Zhong:2007gp,Valcarce:2008dr,Ebert:2011kk,Ortega:2012cx,Yoshida:2015tia,Nagahiro:2016nsx,Wang:2017kfr,Ye:2017yvl,Wang:2018fjm,Gutierrez-Guerrero:2019uwa,Kawakami:2019hpp,Lu:2020ivo,Xiao:2020oif,Chen:2021eyk}, hadronic molecular approaches~\cite{Garcia-Recio:2008rjt,Garcia-Recio:2012lts,Liang:2014eba,An:2017lwg,Montana:2017kjw,Debastiani:2017ewu,Chen:2017xat,Nieves:2017jjx,Huang:2017dwn,Huang:2018bed,Huang:2018wgr,Yu:2018yxl,Nieves:2019jhp,Liang:2020dxr}, hyperfine-interaction analyses~\cite{Copley:1979wj,Karliner:2008sv}, QCD sum rules~\cite{Bagan:1991sg,Neubert:1991sp,Broadhurst:1991fc,Huang:1994zj,Dai:1996yw,Groote:1996em,Colangelo:1998ga,Huang:2000tn,Zhu:2000py,Lee:2000tb,Wang:2003zp,Duraes:2007te,Liu:2007fg,Zhang:2008pm,Aliev:2009jt,Wang:2010it,Zhou:2014ytp,Zhou:2015ywa,Wang:2017zjw,Aliev:2018ube,Aliev:2018vye,Aliev:2018lcs,XuYongJiang:2020cht,Wang:2020pri}, and lattice QCD~\cite{UKQCD:1996ssj,Burch:2008qx,Padmanath:2013bla,Brown:2014ena,Burch:2015pka,Padmanath:2017lng,Can:2019wts,Bahtiyar:2020uuj}. Their production and decay properties have also been studied using the quark-pair-creation model~\cite{Chen:2018orb,Chen:2018vuc,Yang:2018lzg,Liang:2020hbo}, chiral perturbation theory~\cite{Cheng:2006dk,Lu:2014ina,Cheng:2015naa}, and other phenomenological approaches~\cite{Kim:2014qha,Xie:2015zga,Huang:2016ygf,Karliner:2015ema,Chua:2018lfa,Karliner:2018bms,Jia:2019bkr,Wang:2020gkn}. Comprehensive reviews and further discussions can be found in Refs.~\cite{Chen:2016spr,Chen:2022asf,Luo:2025sns,Korner:1994nh,Manohar:2000dt,Bianco:2003vb,Klempt:2009pi,Crede:2013kia,Cheng:2015iom,Cheng:2015rra,Chen:2016qju,Liu:2019zoy,Hosaka:2016pey,Richard:2016eis,Lebed:2016hpi,Esposito:2016noz,Ali:2017jda,Guo:2017jvc,Olsen:2017bmm,Karliner:2017qhf,Guo:2019twa,Brambilla:2019esw,Yang:2020atz,Fang:2021wes,Jin:2021vct,Meng:2022ozq,Liu:2024uxn,Wang:2025dur,Dai:2026fkg}.

The $P$-wave flavor-sextet heavy baryons are of particular interest. Within heavy quark effective theory (HQET), the Pauli principle and angular-momentum couplings allow seven $P$-wave states in each of the $\Sigma_Q$, $\Xi_Q^\prime$, and $\Omega_Q$ sectors ($Q=c,b$). The growing number of experimentally observed structures naturally raises the question of whether the charmed and bottom spectra can be understood within a common framework rooted in heavy-quark symmetry, once finite-heavy-quark-mass effects, strong-decay dynamics, and configuration mixing are properly taken into account. We have previously investigated the masses of $P$-wave flavor-sextet heavy baryons using QCD sum rules within HQET~\cite{Chen:2015kpa,Mao:2015gya}, and studied their strong decay properties using light-cone sum rules~\cite{Chen:2017sci,Yang:2019cvw,Yang:2020zrh}. These studies are based on QCD sum rules~\cite{Shifman:1978bx,Reinders:1984sr}, light-cone sum rules~\cite{Balitsky:1989ry,Braun:1988qv,Chernyak:1990ag,Ball:1998je,Ball:2006wn}, and heavy quark effective theory~\cite{Grinstein:1990mj,Eichten:1989zv,Falk:1990yz}. Related applications of these methods to singly heavy mesons and baryons can be found in Refs.~\cite{Bagan:1991sg,Neubert:1991sp,Broadhurst:1991fc,Huang:1994zj,Dai:1996yw,Colangelo:1998ga,Groote:1996em,Zhu:2000py,Lee:2000tb,Huang:2000tn,Wang:2003zp,Duraes:2007te,Liu:2007fg,Zhou:2014ytp,Zhou:2015ywa}.

A key new development is the recent evidence for the $\Xi_c(2882)^0$ structure~\cite{LHCb:2022vns}. Together with the $\Xi_c(2923)^0$, $\Xi_c(2939)^0$, and $\Xi_c(2965)^0$, it completes a sequence of four narrow $\Xi_c$ structures whose mass-splitting pattern closely parallels that of the four lowest narrow $\Omega_c$ states, $\Omega_c(3000)^0$, $\Omega_c(3050)^0$, $\Omega_c(3066)^0$, and $\Omega_c(3090)^0$. This remarkable correspondence provides substantially stronger phenomenological constraints on their spectroscopic assignments and greatly reduces the ambiguities present in earlier analyses. It therefore strongly motivates a renewed analysis of the $P$-wave flavor-sextet spectrum within a unified framework, extending the study to the less well-resolved $\Sigma_c$ sector and to the corresponding bottom-baryon states.

In the present study, we build upon and substantially extend our previous analyses based on QCD sum rules and light-cone sum rules to develop a unified description of the $P$-wave flavor-sextet charmed and bottom baryons. We first classify these states within HQET and use the experimentally observed mass patterns, in particular the characteristic mass splittings, to constrain their possible spectroscopic assignments. We then systematically investigate their masses and intra-doublet mass splittings using QCD sum rules, study their $S$- and $D$-wave strong decays using light-cone sum rules, and examine configuration mixing among states with the same quantum numbers. By combining these complementary ingredients, we aim to establish a coherent picture of the observed spectrum and to clarify the status of broad, missing, and experimentally unresolved states within a common charm--bottom framework.

This paper is organized as follows. In Sec.~\ref{sec:classification}, we classify the $P$-wave flavor-sextet heavy baryons within HQET and perform phenomenological analyses of the observed charmed and bottom states. In Sec.~\ref{sec:mass}, we investigate their masses and intra-doublet mass splittings using QCD sum rules. In Sec.~\ref{sec:decay}, we study their strong decay properties using light-cone sum rules. In Sec.~\ref{sec:mixing}, we analyze configuration mixing among states with the same quantum numbers. Finally, we summarize and discuss our results in Sec.~\ref{sec:summary}.

\section{Phenomenological analyses}
\label{sec:classification}

In this section, we first classify the $P$-wave flavor-sextet heavy baryons within heavy quark effective theory (HQET), and then analyze the observed charmed and bottom states from a phenomenological perspective. In particular, the experimentally measured mass patterns and mass splittings provide important constraints on their possible spectroscopic assignments. These assignments will be further tested by the mass calculations using QCD sum rules in Sec.~\ref{sec:mass}, the strong decay analyses using light-cone sum rules in Sec.~\ref{sec:decay}, and the study of configuration mixing in Sec.~\ref{sec:mixing}.

In the heavy quark limit, the spin of the heavy quark, $s_Q=1/2$, decouples from the light degrees of freedom. The light degrees of freedom carry a total angular momentum $j_l$, obtained by coupling the spin of the light-quark pair to the orbital angular momenta,
\begin{equation}
j_l=s_l\otimes l_\rho\otimes l_\lambda ,
\end{equation}
where $s_l$ denotes the total spin of the two light quarks. The orbital angular momenta $l_\rho$ and $l_\lambda$ are associated with the two Jacobi coordinates: $l_\rho$ describes the relative orbital motion between the two light quarks, while $l_\lambda$ describes the orbital motion of the heavy quark relative to the center of mass of the light-quark pair. Accordingly, the $\rho$-mode excitation corresponds to $(l_\rho,l_\lambda)=(1,0)$, whereas the $\lambda$-mode excitation corresponds to $(l_\rho,l_\lambda)=(0,1)$. The total angular momentum $J$ of the heavy baryon is then obtained by coupling $j_l$ to the heavy-quark spin,
\begin{equation}
J=
\begin{cases}
1/2, & j_l=0,\\
j_l\pm1/2, & j_l\geq1.
\end{cases}
\end{equation}

\begin{figure*}[hbtp]
\begin{center}
\scalebox{0.65}{\includegraphics{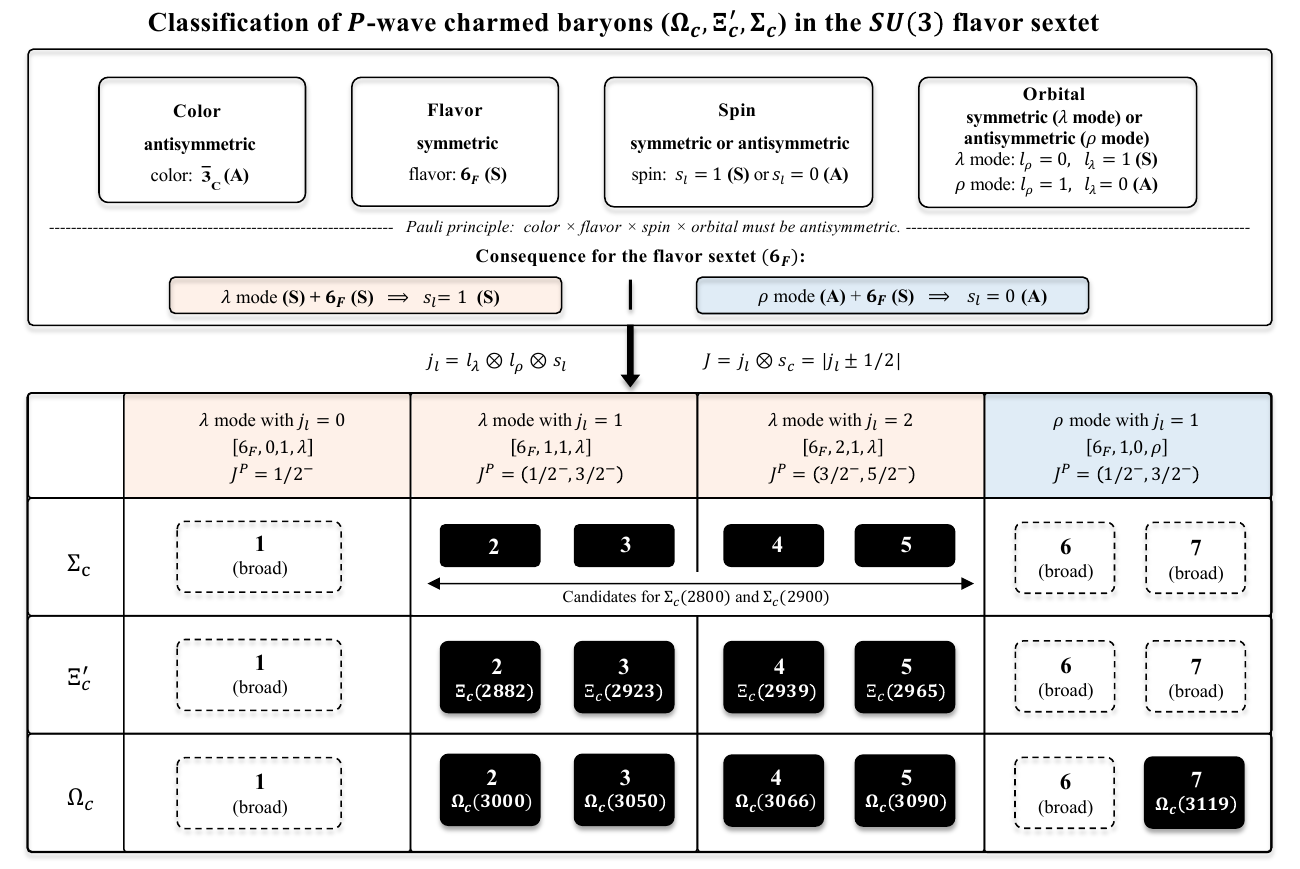}}
\\[2mm]
\scalebox{0.65}{\includegraphics{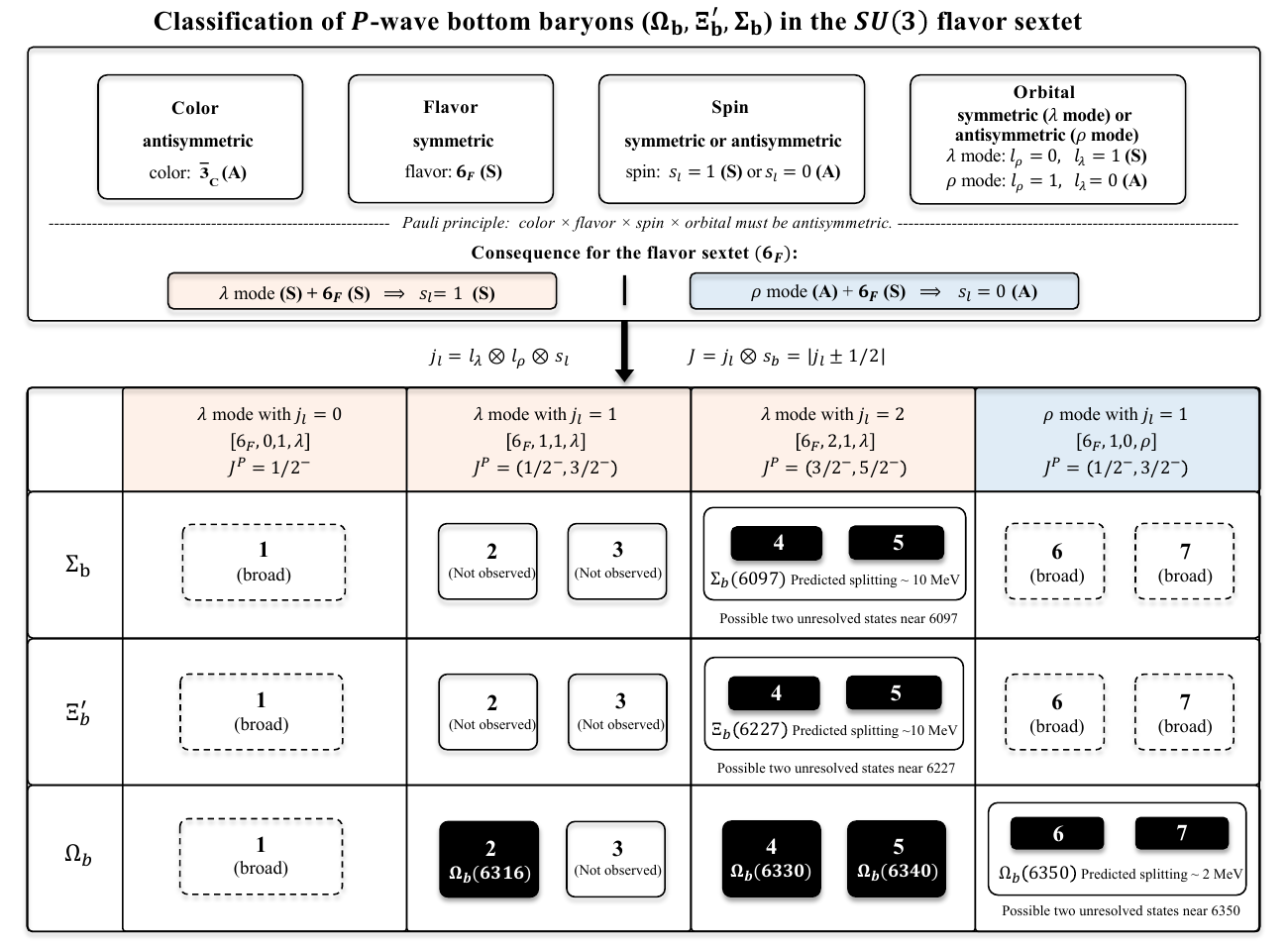}}
\end{center}
\caption{Classification and phenomenological assignments of the $P$-wave flavor-sextet charmed and bottom baryons. The upper and lower panels show the charmed and bottom sectors, respectively. The Pauli principle and angular-momentum couplings organize the seven $P$-wave states in each of the $\Sigma_Q$, $\Xi_Q^\prime$, and $\Omega_Q$ sectors ($Q=c,b$) into four HQET multiplets. Dotted boxes denote relatively broad states, whereas solid boxes denote relatively narrow states. The experimentally observed structures and their possible spectroscopic assignments are highlighted in black. The relevant masses and mass splittings, together with possible experimentally unresolved states, are also indicated.}
\label{fig:classification}
\end{figure*}

For the flavor-sextet heavy baryons, the flavor wave function of the two light quarks is symmetric, whereas their color wave function is antisymmetric. For the $\lambda$-mode excitation, the spatial wave function is symmetric under the interchange of the two light quarks, so the Pauli principle requires their spin wave function to be symmetric, corresponding to $s_l=1$. In contrast, for the $\rho$-mode excitation, the spatial wave function is antisymmetric, and the spin wave function must therefore be antisymmetric, corresponding to $s_l=0$. As illustrated in Fig.~\ref{fig:classification}, the seven $P$-wave states can consequently be classified into four HQET multiplets:
\begin{align}
[{\mathbf 6}_F,0,1,\lambda] &: \qquad
J^P=1/2^- ,
\nonumber\\
[{\mathbf 6}_F,1,1,\lambda] &: \qquad
J^P=(1/2^-,3/2^-) ,
\nonumber\\
[{\mathbf 6}_F,2,1,\lambda] &: \qquad
J^P=(3/2^-,5/2^-) ,
\nonumber\\
[{\mathbf 6}_F,1,0,\rho] &: \qquad
J^P=(1/2^-,3/2^-) .
\label{eq:hqet-classification}
\end{align}
Here we use the notation $[{\mathbf 6}_F,j_l,s_l,\rho/\lambda]$ to specify the flavor representation, the total angular momentum of the light degrees of freedom, the spin of the two light quarks, and the orbital excitation mode, respectively. As an example, the seven $P$-wave $\Omega_c$ states are distributed among these four HQET multiplets as
\begin{align}
[\Omega_c,0,1,\lambda] &: \qquad
\Omega_c(1/2^-),
\nonumber\\
[\Omega_c,1,1,\lambda] &: \qquad
\Omega_c(1/2^-),\,
\Omega_c(3/2^-),
\nonumber\\
[\Omega_c,2,1,\lambda] &: \qquad
\Omega_c(3/2^-),\,
\Omega_c(5/2^-),
\nonumber\\
[\Omega_c,1,0,\rho] &: \qquad
\Omega_c(1/2^-),\,
\Omega_c(3/2^-).
\label{eq:omegac-classification}
\end{align}

In the strict heavy quark limit, the two members of each $j_l\neq0$ multiplet are degenerate. Their mass splittings, as well as configuration mixing between states with the same $J^P$, arise from finite-heavy-quark-mass effects. Since the resulting mass splittings are generally much smaller than the overall excitation energies, the characteristic mass-splitting patterns observed experimentally provide particularly useful information for identifying the underlying HQET multiplets. We shall therefore use these experimental mass patterns and splittings as important phenomenological constraints in the analyses below.

\subsection{Charmed baryons}
\label{subsec:charm}

The experimentally observed excited charmed baryons provide a natural starting point for identifying their possible HQET assignments. As summarized in the upper panel of Fig.~\ref{fig:classification}, four excited $\Xi_c$ structures, $\Xi_c(2882)$, $\Xi_c(2923)$, $\Xi_c(2939)$, and $\Xi_c(2965)$, have been observed in the relevant mass region, while five narrow $\Omega_c$ structures, $\Omega_c(3000)$, $\Omega_c(3050)$, $\Omega_c(3066)$, $\Omega_c(3090)$, and $\Omega_c(3119)$, have been established experimentally.

A particularly striking feature is the similarity between the mass patterns of the four $\Xi_c$ structures and the four lowest narrow $\Omega_c$ structures. Using their nominal masses, the successive mass splittings between adjacent structures are approximately
\begin{align}
\Delta M_{\Xi_c}
&\simeq (41,\,16,\,26)~{\rm MeV},
\nonumber\\
\Delta M_{\Omega_c}
&\simeq (50,\,16,\,24)~{\rm MeV}.
\label{eq:charm-splittings}
\end{align}
This close correspondence strongly suggests that the two sequences may share the same underlying HQET structure and therefore provides an important constraint on their spectroscopic assignments.

Our previous light-cone sum rules analyses further showed that the four $\lambda$-mode states belonging to the $[{\mathbf 6}_F,1,1,\lambda]$ and $[{\mathbf 6}_F,2,1,\lambda]$ multiplets are generally relatively narrow in the $\Xi_c^\prime$ and $\Omega_c$ sectors, and are therefore more likely to be observed experimentally~\cite{Chen:2017sci,Yang:2019cvw,Yang:2020zrh,Luo:2026elv}. In contrast, the $\Xi_c^\prime(1/2^-)$ and $\Omega_c(1/2^-)$ states in the $[{\mathbf 6}_F,0,1,\lambda]$ multiplet and the two $\Xi_c^\prime(1/2^-,3/2^-)$ states in the $[{\mathbf 6}_F,1,0,\rho]$ multiplet are predicted to be considerably broader and hence more difficult to identify experimentally.

Based on the characteristic mass-splitting pattern and the expected decay-width hierarchy, we associate the four narrow $\Xi_c$ structures with the four $\lambda$-mode states belonging to the $[{\mathbf 6}_F,1,1,\lambda]$ and $[{\mathbf 6}_F,2,1,\lambda]$ multiplets:
\begin{align}
\Xi_c(2882)
&\leftrightarrow
[\Xi_c^\prime(1/2^-),1,1,\lambda],
\nonumber\\
\Xi_c(2923)
&\leftrightarrow
[\Xi_c^\prime(3/2^-),1,1,\lambda],
\nonumber\\
\Xi_c(2939)
&\leftrightarrow
[\Xi_c^\prime(3/2^-),2,1,\lambda],
\nonumber\\
\Xi_c(2965)
&\leftrightarrow
[\Xi_c^\prime(5/2^-),2,1,\lambda].
\label{eq:xic-assignments}
\end{align}
The close correspondence between the $\Xi_c$ and $\Omega_c$ mass patterns then suggests the analogous assignments for the four lowest narrow $\Omega_c$ structures:
\begin{align}
\Omega_c(3000)
&\leftrightarrow
[\Omega_c(1/2^-),1,1,\lambda],
\nonumber\\
\Omega_c(3050)
&\leftrightarrow
[\Omega_c(3/2^-),1,1,\lambda],
\nonumber\\
\Omega_c(3066)
&\leftrightarrow
[\Omega_c(3/2^-),2,1,\lambda],
\nonumber\\
\Omega_c(3090)
&\leftrightarrow
[\Omega_c(5/2^-),2,1,\lambda].
\label{eq:omegac-assignments}
\end{align}
These assignments should be understood as specifying the dominant HQET components of the observed structures. In particular, the two $J^P=3/2^-$ states in each flavor sector carry the same overall quantum numbers and can therefore mix through finite-heavy-quark-mass effects. Such configuration mixing will be studied explicitly in Sec.~\ref{sec:mixing}.

The additional narrow $\Omega_c(3119)$ structure is particularly intriguing. Once the four lower $\Omega_c$ structures are associated with the two $\lambda$-mode doublets discussed above, the $\Omega_c(3119)$ cannot be naturally accommodated within either of them. Among the remaining $P$-wave configurations, we tentatively associate it with the $J^P=3/2^-$ member of the $\rho$-mode $[{\mathbf 6}_F,1,0,\rho]$ doublet:
\begin{equation}
\Omega_c(3119)
\leftrightarrow
[\Omega_c(3/2^-),1,0,\rho].
\label{eq:omegac3119-assignment}
\end{equation}
Both its $J^P=1/2^-$ partner, $[\Omega_c(1/2^-),1,0,\rho]$, and the $[\Omega_c(1/2^-),0,1,\lambda]$ state are expected to be considerably broader and hence more difficult to identify experimentally. The different experimental visibility of these states will be further examined through their strong decay properties in Sec.~\ref{sec:decay}.

The situation in the $\Sigma_c$ sector is less transparent. The corresponding $P$-wave states are generally expected to be broader than their $\Xi_c^\prime$ and $\Omega_c$ counterparts, so that several nearby states may overlap in the experimental mass spectrum. In particular, we do not necessarily interpret the observed $\Sigma_c(2800)$ and $\Sigma_c(2900)$ structures as individual resonances. Instead, as illustrated in the upper panel of Fig.~\ref{fig:classification}, these structures may receive contributions from several $P$-wave $\Sigma_c$ states, especially the four $\lambda$-mode states belonging to the $[{\mathbf 6}_F,1,1,\lambda]$ and $[{\mathbf 6}_F,2,1,\lambda]$ multiplets. This possibility will be examined below by combining the mass spectrum obtained from QCD sum rules with the predicted strong decay widths.

\subsection{Bottom baryons}
\label{subsec:bottom}

The experimental situation in the bottom sector is considerably different from that in the charm sector. As discussed above, one excited structure has been observed in each of the $\Sigma_b$ and $\Xi_b^\prime$ sectors, namely $\Sigma_b(6097)$~\cite{LHCb:2018haf} and $\Xi_b(6227)$~\cite{LHCb:2018vuc,LHCb:2020xpu}, while four narrow $\Omega_b$ structures, $\Omega_b(6316)$, $\Omega_b(6330)$, $\Omega_b(6340)$, and $\Omega_b(6350)$, have been established experimentally. Since spin-dependent interactions are suppressed in the bottom sector, the mass splittings within an HQET doublet are generally smaller than those in the corresponding charm sector. Consequently, two nearby states may remain experimentally unresolved if their decay widths are comparable to or larger than their mass splitting.

We first consider the $\Sigma_b$ and $\Xi_b^\prime$ sectors. We associate the observed $\Sigma_b(6097)$ and $\Xi_b(6227)$ structures predominantly with the $[{\mathbf 6}_F,2,1,\lambda]$ doublet, but do not necessarily identify each observed structure with a single member of the doublet. Instead, each structure may represent an unresolved overlap of the $J^P=3/2^-$ and $5/2^-$ states:
\begin{align}
\Sigma_b(6097):\qquad
&[\Sigma_b(3/2^-),2,1,\lambda],
\nonumber\\
&[\Sigma_b(5/2^-),2,1,\lambda],
\label{eq:sigmab-assignment}
\end{align}
and
\begin{align}
\Xi_b(6227):\qquad
&[\Xi_b^\prime(3/2^-),2,1,\lambda],
\nonumber\\
&[\Xi_b^\prime(5/2^-),2,1,\lambda].
\label{eq:xib-assignment}
\end{align}
The lower $[{\mathbf 6}_F,1,1,\lambda]$ doublets in the $\Sigma_b$ and $\Xi_b^\prime$ sectors have not yet been identified experimentally. As will be discussed in Sec.~\ref{sec:decay}, their decay patterns differ from those of the $[{\mathbf 6}_F,2,1,\lambda]$ states, and they may therefore be more accessible in other decay channels.

The $\Omega_b$ sector presents a richer situation. Motivated by the unified charm--bottom picture, we propose that the observed $\Omega_b$ spectrum may accommodate up to six narrow $P$-wave states:
\begin{align}
\Omega_b(6316)
\leftrightarrow
&\,[\Omega_b(1/2^-),1,1,\lambda],
\nonumber\\
&\,[\Omega_b(3/2^-),1,1,\lambda],
\nonumber\\
\Omega_b(6330)
\leftrightarrow
&\,[\Omega_b(3/2^-),2,1,\lambda],
\nonumber\\
\Omega_b(6340)
\leftrightarrow
&\,[\Omega_b(5/2^-),2,1,\lambda],
\label{eq:omegab-lambda-assignment}
\end{align}
together with
\begin{align}
\Omega_b(6350):\qquad
&[\Omega_b(1/2^-),1,0,\rho],
\nonumber\\
&[\Omega_b(3/2^-),1,0,\rho].
\label{eq:omegab-rho-assignment}
\end{align}
The missing $[\Omega_b(3/2^-),1,1,\lambda]$ configuration may lie in the vicinity of $\Omega_b(6316)$, although its precise mass and experimental visibility remain uncertain. The two members of the $\rho$-mode $[{\mathbf 6}_F,1,0,\rho]$ doublet are expected to be nearly degenerate in the bottom sector and may likewise remain unresolved within the $\Omega_b(6350)$ structure.

\section{Mass spectrum}
\label{sec:mass}

\begin{table*}[hbtp]
\begin{center}
\renewcommand{\arraystretch}{1.3}
\caption{QCD sum rules parameters and mass predictions for the $P$-wave flavor-sextet charmed and bottom baryons. The isospin factors are explicitly taken into account in the decay constants, such that $f_{\Sigma_c^{++}}=f_{\Sigma_c^0}=\sqrt{2}f_{\Sigma_c^+}$, $f_{\Xi_c^{\prime+}}=f_{\Xi_c^{\prime0}}$, $f_{\Sigma_b^+}=f_{\Sigma_b^-}=\sqrt{2}f_{\Sigma_b^0}$, and $f_{\Xi_b^{\prime0}}=f_{\Xi_b^{\prime-}}$. Relevant central masses are quoted to
the nearest MeV to facilitate comparison and display the
splittings, without implying such precision for the absolute
masses.}
\begin{tabular}{c | c | c | c | c | c c | c | c}
\hline\hline
\multirow{2}{*}{Multiplets} & \multirow{2}{*}{~B~} & $\omega_c$ & ~~~Working region~~~ & ~~~~~~~$\overline{\Lambda}$~~~~~~~ & ~~~Baryons~~~ & ~~~Mass~~~ & ~Splitting~ & $f$
\\                                              &  & (GeV)      & (GeV)                & (GeV)                              & ($J^P$)       & (GeV)      & (MeV)        & (GeV$^{4}$)
\\ \hline\hline
\multirow{3}{*}{$[\mathbf{6}_F, 0, 1, \lambda]$} & $\Sigma_c$ & $1.35$ & $T=0.27$ & $1.10 \pm 0.05$ & $\Sigma_c(1/2^-)$ & $2.829 \pm 0.059$ & -- & $0.045 \pm 0.008~(\Sigma^0_c(1/2^-))$
\\ \cline{2-9}
                                                 & $\Xi^\prime_c$ & $1.54$ & $0.27< T < 0.28$ & $1.21 \pm 0.08$ & $\Xi^\prime_c(1/2^-)$ & $2.885 \pm 0.124$ & -- & $0.039 \pm 0.010~(\Xi^{\prime0}_c(1/2^-))$
\\ \cline{2-9}
                                                 & $\Omega_c$ & 1.76 & $0.27< T < 0.30$ & $1.36 \pm 0.09$ & $\Omega_c(1/2^-)$ & $3.005 \pm 0.190$ & -- & $0.078 \pm 0.021~(\Omega^0_c(1/2^-))$
\\ \hline
\multirow{6}{*}{$[\mathbf{6}_F, 1, 1, \lambda]$}
& \multirow{2}{*}{$\Sigma_c$} & \multirow{2}{*}{1.78} & \multirow{2}{*}{$0.32<T<0.34$} & \multirow{2}{*}{$1.10 \pm 0.12$} & $\Sigma_c(1/2^-)$ & $2.832 \pm 0.172$ & \multirow{2}{*}{$36 \pm15 $} & $0.053 \pm0.013 ~(\Sigma^0_c(1/2^-))$
\\ \cline{6-7}\cline{9-9}
& & & & & $\Sigma_c(3/2^-)$ & $2.868 \pm0.167 $ & &$0.025\pm 0.006~(\Sigma^0_c(3/2^-))$
\\ \cline{2-9}
& \multirow{2}{*}{$\Xi^\prime_c$} & \multirow{2}{*}{1.70} & \multirow{2}{*}{$T=0.34$} & \multirow{2}{*}{$1.12 \pm 0.09$} & $\Xi^\prime_c(1/2^-)$ & $2.890 \pm 0.132$ & \multirow{2}{*}{$39 \pm 15$} & $0.039 \pm 0.008~(\Xi^{\prime0}_c(1/2^-))$
\\ \cline{6-7}\cline{9-9}
& & & & & $\Xi^\prime_c(3/2^-)$ & $2.929 \pm0.128$ & &$ 0.018\pm 0.004~(\Xi^{\prime0}_c(3/2^-))$
\\ \cline{2-9}
& \multirow{2}{*}{$\Omega_c$} & \multirow{2}{*}{1.70} & \multirow{2}{*}{$T=0.35$} & \multirow{2}{*}{$1.21 \pm 0.08$} & $\Omega_c(1/2^-)$ & $3.019 \pm0.106 $ & \multirow{2}{*}{$36\pm 14$} & $0.067 \pm 0.012~(\Omega^0_c(1/2^-))$
\\ \cline{6-7}\cline{9-9}
& & & & & $\Omega_c(3/2^-)$ & $3.055\pm0.102 $ & &$ 0.032\pm 0.006~(\Omega^0_c(3/2^-))$
\\ \hline
\multirow{6}{*}{$[\mathbf{6}_F, 2, 1, \lambda]$}
& \multirow{2}{*}{$\Sigma_c$} & \multirow{2}{*}{1.60}& \multirow{2}{*}{$0.27< T < 0.30$} & \multirow{2}{*}{$1.15 \pm0.15 $} & $\Sigma_c(3/2^-)$ & $2.873 \pm0.228 $ & \multirow{2}{*}{$74 \pm32 $} & $0.065 \pm 0.023~(\Sigma^0_c(3/2^-))$
\\ \cline{6-7}\cline{9-9}
& & & & & $\Sigma_c(5/2^-)$ & $2.947 \pm0.217 $ & &$0.028\pm0.010 ~(\Sigma^0_c(5/2^-))$
\\ \cline{2-9}
 & \multirow{2}{*}{$\Xi^\prime_c$} & \multirow{2}{*}{1.69}& \multirow{2}{*}{$0.27< T < 0.31$} & \multirow{2}{*}{$1.22 \pm0.15 $} & $\Xi^\prime_c(3/2^-)$ & $2.932 \pm0.238 $ & \multirow{2}{*}{$68 \pm30 $} & $0.054 \pm0.019 ~(\Xi^{\prime0}_c(3/2^-))$
\\ \cline{6-7}\cline{9-9}
& & & & & $\Xi^\prime_c(5/2^-)$ & $3.000 \pm0.220 $ & &$ 0.023\pm0.008  ~(\Xi^{\prime0}_c(5/2^-))$
\\ \cline{2-9}
& \multirow{2}{*}{$\Omega_c$} & \multirow{2}{*}{1.84} & \multirow{2}{*}{$0.26< T < 0.33$} & \multirow{2}{*}{$1.34 \pm 0.14$} & $\Omega_c(3/2^-)$ & $3.059\pm0.230 $ & \multirow{2}{*}{$60 \pm27 $} & $0.100 \pm 0.029~(\Omega^0_c(3/2^-))$
\\ \cline{6-7}\cline{9-9}
& & & & & $\Omega_c(5/2^-)$ & $3.119 \pm 0.217$ & &$ 0.042\pm0.012 ~(\Omega^0_c(5/2^-))$
\\\hline
\multirow{6}{*}{$[\mathbf{6}_F, 1, 0, \rho]$}
& \multirow{2}{*}{$\Sigma_c$} & \multirow{2}{*}{1.85} & \multirow{2}{*}{$0.26< T < 0.34$} & \multirow{2}{*}{$1.34 \pm 0.13$} & $\Sigma_c(1/2^-)$ & $2.890 \pm 0.149$ & \multirow{2}{*}{$13 \pm 6$} & $0.082 \pm 0.024~(\Sigma^0_c(1/2^-))$
\\ \cline{6-7}\cline{9-9}
& & & & & $\Sigma_c(3/2^-)$ & $2.903 \pm 0.147$ & &$0.039 \pm 0.011~(\Sigma^0_c(3/2^-))$
\\ \cline{2-9}
& \multirow{2}{*}{$\Xi^\prime_c$} & \multirow{2}{*}{1.95} & \multirow{2}{*}{$0.26< T < 0.35$} & \multirow{2}{*}{$1.42 \pm 0.11$} & $\Xi^\prime_c(1/2^-)$ & $2.976 \pm 0.140$ & \multirow{2}{*}{$12 \pm 5$} & $0.069 \pm 0.018~(\Xi^{\prime0}_c(1/2^-))$
\\ \cline{6-7}\cline{9-9}
& & & & & $\Xi^\prime_c(3/2^-)$ & $2.989 \pm 0.140$ & &$0.032 \pm 0.008~(\Xi^{\prime0}_c(3/2^-))$
\\ \cline{2-9}
& \multirow{2}{*}{$\Omega_c$} & \multirow{2}{*}{2.08} & \multirow{2}{*}{$0.26< T < 0.37$} & \multirow{2}{*}{$1.55\pm0.10 $} & $\Omega_c(1/2^-)$ & $3.106 \pm0.159 $ & \multirow{2}{*}{$12\pm5$} & $0.124\pm0.027 ~(\Omega^0_c(1/2^-))$
\\ \cline{6-7}\cline{9-9}
& & & & & $\Omega_c(3/2^-)$ & $3.117\pm0.160 $ & &$ 0.058\pm0.013 ~(\Omega^0_c(3/2^-))$
\\ \hline \hline
\multirow{3}{*}{$[\mathbf{6}_F, 0, 1, \lambda]$} & $\Sigma_b$ & $1.70$ & $0.26< T < 0.32$ & $1.25 \pm 0.10$ & $\Sigma_b(1/2^-)$ & $6.051 \pm 0.109$ & -- & $0.077 \pm 0.020~(\Sigma^-_b(1/2^-))$
\\ \cline{2-9}
                                                 & $\Xi^\prime_b$ & $1.83$ & $0.27< T < 0.33$ & $1.38 \pm 0.10$ & $\Xi^\prime_b(1/2^-)$ & $6.180 \pm 0.117$ & -- & $0.067 \pm 0.017~(\Xi^{\prime-}_b(1/2^-))$
\\ \cline{2-9}
                                                 & $\Omega_b$ & 1.96 & $0.27< T < 0.33$ & $1.51 \pm 0.09$ & $\Omega_b(1/2^-)$ & $6.302 \pm 0.119$ & -- & $0.116 \pm 0.028~(\Omega^-_b(1/2^-))$
\\ \hline
\multirow{6}{*}{$[\mathbf{6}_F, 1, 1, \lambda]$}
& \multirow{2}{*}{$\Sigma_b$} & \multirow{2}{*}{1.95} & \multirow{2}{*}{$0.29< T < 0.36$} & \multirow{2}{*}{$1.26\pm0.11 $} & $\Sigma_b(1/2^-)$ & $6.075\pm0.128$ & \multirow{2}{*}{$6 \pm 3$} & $0.077 \pm 0.018~(\Sigma^-_b(1/2^-))$
\\ \cline{6-7}\cline{9-9}
& & & & & $\Sigma_b(3/2^-)$ & $6.081 \pm0.129$ & &$0.036 \pm0.008 ~(\Sigma^-_b(3/2^-))$
\\ \cline{2-9}
& \multirow{2}{*}{$\Xi^\prime_b$} & \multirow{2}{*}{1.96} & \multirow{2}{*}{$0.35< T < 0.37$} & \multirow{2}{*}{$1.37 \pm0.09 $} & $\Xi^\prime_b(1/2^-)$ & $6.206\pm0.111 $ & \multirow{2}{*}{$7 \pm2 $} & $0.068 \pm 0.013~(\Xi^{\prime-}_b(1/2^-))$
\\ \cline{6-7}\cline{9-9}
& & & & & $\Xi^\prime_b(3/2^-)$ & $6.213 \pm 0.111$ & &$0.032 \pm0.006 ~(\Xi^{\prime-}_b(3/2^-))$
\\ \cline{2-9}
& \multirow{2}{*}{$\Omega_b$} & \multirow{2}{*}{1.97} & \multirow{2}{*}{$ T =0.38$} & \multirow{2}{*}{$1.47 \pm 0.08$} & $\Omega_b(1/2^-)$ & $6.315 \pm 0.092$ & \multirow{2}{*}{$7 \pm 2$} & $0.116 \pm0.020~(\Omega^-_b(1/2^-))$
\\ \cline{6-7}\cline{9-9}
& & & & & $\Omega_b(3/2^-)$ & $6.322 \pm 0.091$ & &$0.055 \pm0.009 ~(\Omega^-_b(3/2^-))$
\\ \hline
\multirow{6}{*}{$[\mathbf{6}_F, 2, 1, \lambda]$}
& \multirow{2}{*}{$\Sigma_b$} & \multirow{2}{*}{1.82} & \multirow{2}{*}{$0.27< T < 0.33$} & \multirow{2}{*}{$1.28 \pm 0.16$} & $\Sigma_b(3/2^-)$ & $6.089\pm 0.182$ & \multirow{2}{*}{$12 \pm 6$} & $0.098 \pm 0.034~(\Sigma^-_b(3/2^-))$
\\ \cline{6-7}\cline{9-9}
& & & & & $\Sigma_b(5/2^-)$ & $6.101 \pm0.180 $ & &$0.059 \pm 0.020~(\Sigma^-_b(5/2^-))$
\\ \cline{2-9}
 & \multirow{2}{*}{$\Xi^\prime_b$} & \multirow{2}{*}{1.95} & \multirow{2}{*}{$0.26< T < 0.35$} & \multirow{2}{*}{$1.40 \pm 0.13$} & $\Xi^\prime_b(3/2^-)$ & $6.218 \pm0.157$ & \multirow{2}{*}{$11 \pm 5$} & $0.089 \pm 0.026~(\Xi^{\prime-}_b(3/2^-))$
\\ \cline{6-7}\cline{9-9}
& & & & & $\Xi^\prime_b(5/2^-)$ & $6.230 \pm0.155$ & &$0.053 \pm 0.016~(\Xi^{\prime-}_b(5/2^-))$
\\ \cline{2-9}
& \multirow{2}{*}{$\Omega_b$} & \multirow{2}{*}{2.06} & \multirow{2}{*}{$0.26< T < 0.36$} & \multirow{2}{*}{$1.51 \pm0.10$} & $\Omega_b(3/2^-)$ & $6.331\pm0.126 $ & \multirow{2}{*}{$10 \pm 5$} & $0.156 \pm0.036 ~(\Omega^-_b(3/2^-))$
\\ \cline{6-7}\cline{9-9}
& & & & & $\Omega_b(5/2^-)$ & $6.341 \pm0.125 $ & &$0.093 \pm0.022 ~(\Omega^-_b(5/2^-))$
\\\hline
\multirow{6}{*}{$[\mathbf{6}_F, 1, 0, \rho]$}
& \multirow{2}{*}{$\Sigma_b$} & \multirow{2}{*}{1.90} & \multirow{2}{*}{$0.26< T < 0.35$} & \multirow{2}{*}{$1.36 \pm0.12$} & $\Sigma_b(1/2^-)$ & $6.108 \pm 0.128$ & \multirow{2}{*}{$3 \pm 1$} & $0.090 \pm 0.025~(\Sigma^-_b(1/2^-))$
\\ \cline{6-7}\cline{9-9}
& & & & & $\Sigma_b(3/2^-)$ & $6.111 \pm 0.127$ & &$0.043 \pm 0.012~(\Sigma^-_b(3/2^-))$
\\ \cline{2-9}
& \multirow{2}{*}{$\Xi^\prime_b$} & \multirow{2}{*}{2.04} & \multirow{2}{*}{$0.26< T < 0.36$} & \multirow{2}{*}{$1.49 \pm0.10 $} & $\Xi^\prime_b(1/2^-)$ & $6.240 \pm0.116 $ & \multirow{2}{*}{$2 \pm 1$} & $0.081 \pm 0.019~(\Xi^{\prime-}_b(1/2^-))$
\\ \cline{6-7}\cline{9-9}
& & & & & $\Xi^\prime_b(3/2^-)$ & $6.242 \pm 0.116$ & &$0.038 \pm 0.009~(\Xi^{\prime-}_b(3/2^-))$
\\ \cline{2-9}
& \multirow{2}{*}{$\Omega_b$} & \multirow{2}{*}{2.16} & \multirow{2}{*}{$0.26< T < 0.38$} & \multirow{2}{*}{$1.61 \pm 0.11$} & $\Omega_b(1/2^-)$ & $6.351 \pm0.128 $ & \multirow{2}{*}{$2 \pm 1$} & $0.141 \pm 0.032~(\Omega^-_b(1/2^-))$
\\ \cline{6-7}\cline{9-9}
& & & & & $\Omega_b(3/2^-)$ & $6.353 \pm 0.129$ & &$0.067\pm 0.015~(\Omega^-_b(3/2^-))$
\\ \hline \hline
\end{tabular}
\label{tab:mass}
\end{center}
\end{table*}

In this section, we investigate the masses and intra-doublet mass splittings of the $P$-wave flavor-sextet heavy baryons using QCD sum rules within heavy quark effective theory (HQET). Our analyses follow the framework developed in Refs.~\cite{Chen:2015kpa,Mao:2015gya}, with updated numerical inputs and a unified treatment of the charmed and bottom sectors. As a representative example, we consider the $\Omega_b(3/2^-)$ baryon belonging to the $[{\mathbf 6}_F,2,1,\lambda]$ multiplet. The same state will also be used in Sec.~\ref{sec:decay} to illustrate the light-cone sum rules analyses of strong decays.

The interpolating current for the $[\Omega_b(3/2^-),2,1,\lambda]$ baryon is
\begin{eqnarray}
J^\alpha_{\Omega_b(3/2^-)}
&=& i\epsilon_{abc}\Big([\mathcal{D}_t^\mu s^{aT}]C\gamma_t^\nu s^b+s^{aT}C\gamma_t^\nu[\mathcal{D}_t^\mu s^b]\Big)
\nonumber\\
&\times& \left(g_t^{\alpha\mu}\gamma_t^\nu\gamma_5+g_t^{\alpha\nu}\gamma_t^\mu\gamma_5-\frac{2}{3}g_t^{\mu\nu}\gamma_t^\alpha\gamma_5\right)h_v^c ,
\nonumber\\
\label{eq:current-omegab32}
\end{eqnarray}
where $a\cdots c$ are color indices, $C$ is the charge-conjugation operator, $\mathcal{D}_t^\mu=\mathcal{D}^\mu-(v\cdot\mathcal{D})v^\mu$, $\gamma_t^\mu=\gamma^\mu-v\!\!\!/\;v^\mu$, and $g_t^{\mu\nu}=g^{\mu\nu}-v^\mu v^\nu$. The interpolating currents for the other $P$-wave flavor-sextet heavy baryons can be constructed in the same manner and have been given in Refs.~\cite{Chen:2015kpa,Mao:2015gya}. The coupling of this current to the $\Omega_b(3/2^-)$ baryon is defined by
\begin{equation}
\langle 0|J^\alpha_{\Omega_b(3/2^-)}|\Omega_b(3/2^-),2,1,\lambda\rangle
=
f \times u^\alpha ,
\label{eq:mass-coupling}
\end{equation}
where $f$ is the decay constant and $u^\alpha$ is the Rarita--Schwinger spinor.

We construct the two-point correlation function
\begin{eqnarray}
\Pi^{\alpha\beta}(\omega)
&=&i\int d^4xe^{ik\cdot x}\langle0|T[J^\alpha_{\Omega_b(3/2^-)}(x)\bar J^\beta_{\Omega_b(3/2^-)}(0)]|0\rangle 
\nonumber \\ &=& {1 + v\!\!\!\slash \over 2} \, g_t^{\alpha \beta} \, \Pi(\omega),
\label{eq:mass-correlator}
\end{eqnarray}
where $\omega=v\cdot k$ denotes the off-shell energy. The invariant function $\Pi(\omega)$ is then used to perform the QCD sum rule analysis. At the hadronic level, its lowest-lying pole contribution is written as
\begin{equation}
\Pi(\omega)=\frac{f^2}{\overline{\Lambda}-\omega}+\textrm{higher resonances},
\label{eq:mass-pole}
\end{equation}
where
\begin{equation}
\overline{\Lambda}\equiv\lim_{m_b\to\infty}(m_{\Omega_b}-m_b)
\label{eq:barLambda}
\end{equation}
is the residual mass of the $\Omega_b(3/2^-)$ baryon in the heavy quark limit.

At the quark-gluon level, the correlation function is evaluated using the operator product expansion (OPE). After performing the Borel transformation and invoking quark--hadron duality to subtract the contributions from higher resonances and continuum states, we obtain the Borel-transformed sum rule
\begin{align}
&\Pi(\omega_c, T) = f^2e^{-\overline{\Lambda}/T},\nonumber\\
&= \int_{2m_s}^{\omega_c} \left[\frac{4\omega^{7}}{63\pi^4}-\frac{2m_s^2\omega^{5}}{3\pi^4}+\frac{5m_s^4\omega^{3}}{3\pi^4}\right]
e^{-\omega/T} \, d\omega \nonumber\\
&\quad -\frac{5\langle g^2 GG \rangle T^4}{24\pi^4}-\frac{5m_s^3\langle\bar{s}s\rangle T^2}{3\pi^2}+\frac{5\langle g^2 GG \rangle m_s^2 T^2}{96\pi^4}\nonumber\\
&\quad - \frac{5\langle g_s \bar{s} \sigma G s \rangle \langle\bar{s}s\rangle}{18} -\frac{5\langle g^2 GG \rangle m_s\langle\bar{s}s\rangle}{576\pi^2}\nonumber\\
&\quad - \frac{5\langle g_s \bar{s} \sigma G s \rangle^2 }{288T^2}\, ,
\label{eq:mass-sumrule-general}
\end{align}
where $\omega_c$ is the threshold value and $T$ is the Borel mass. The residual mass and decay constant can then be extracted from this sum rule as
\begin{eqnarray}
\overline{\Lambda}(\omega_c,T)
&=&\frac{1}{\Pi(\omega_c,T)}\frac{\partial\Pi(\omega_c,T)}{\partial(-1/T)},
\label{eq:barLambda-extraction}
\\
f^2(\omega_c,T)
&=&\Pi(\omega_c,T) \times e^{\overline{\Lambda}(\omega_c,T)/T}.
\label{eq:f-extraction}
\end{eqnarray}
In the numerical analyses, we work at the renormalization scale $\mu=2~{\rm GeV}$ for the bottom sector and $\mu=1~{\rm GeV}$ for the charm sector. The threshold value $\omega_c$ and the Borel mass $T$ are constrained by the convergence of the OPE, the pole contribution, and the stability of the extracted results. The working regions are chosen such that the higher-dimensional power corrections remain sufficiently small, the pole contribution is sufficiently large, and the extracted quantities show reasonable stability against variations of $\omega_c$ and $T$.

To calculate the mass splittings between the two members of an HQET doublet, we further include the ${\mathcal O}(1/m_Q)$ corrections~\cite{Dai:1996qx,Dai:2003yg}. The HQET effective Lagrangian up to this order is
\begin{equation}
\mathcal{L}_{\rm eff}
=
\bar h_v\,iv\cdot D\,h_v
+\frac{1}{2m_Q}\mathcal{K}
+\frac{1}{2m_Q}\mathcal{S},
\label{eq:hqet-lagrangian}
\end{equation}
where
\begin{align}
\mathcal{K}
&=
\bar h_v(iD_\perp)^2h_v,
\\
\mathcal{S}
&=
\frac{g_s}{2}
C_{\rm mag}\left(\frac{m_Q}{\mu}\right)
\bar h_v\sigma_{\mu\nu}G^{\mu\nu}h_v,
\end{align}
denote the kinetic-energy and chromomagnetic operators, respectively. The Wilson coefficient is given by
\begin{equation}
C_{\rm mag}\left(\frac{m_Q}{\mu}\right)
=
\left[
\frac{\alpha_s(m_Q)}
{\alpha_s(\mu)}
\right]^{3/\beta_0},
\qquad
\beta_0=11-\frac{2}{3}n_f .
\label{eq:Cmag}
\end{equation}
The corresponding hadronic matrix elements are parametrized as
\begin{align}
K
&\equiv
\langle B(v)|\bar h_v(iD_\perp)^2h_v|B(v)\rangle,
\nonumber\\
d_M\Sigma
&\equiv
\left\langle B(v)\left|
\frac{g_s}{2}\bar h_v\sigma_{\mu\nu}G^{\mu\nu}h_v
\right|B(v)\right\rangle,
\label{eq:K-Sigma}
\end{align}
where the coefficient $d_M=d_{J,j_l}$ is given by
\begin{align}
d_{j_l-1/2,j_l}
&=2j_l+2,
\nonumber\\
d_{j_l+1/2,j_l}
&=-2j_l.
\label{eq:dM}
\end{align}

To determine $K$ and $\Sigma$, we consider the three-point correlation functions
\begin{eqnarray}
\delta_O\Pi^{\alpha\beta}(\omega,\omega^\prime)
&=&i^2\int d^4x\,d^4y\,e^{ik\cdot x-ik^\prime\cdot y}
\label{eq:mass-three-point}
\\
&\times&\langle0|T[J^\alpha_{\Omega_b(3/2^-)}(x)O(0)\bar J^\beta_{\Omega_b(3/2^-)}(y)]|0\rangle
\nonumber\\
&=& {1 + v\!\!\!\slash \over 2} \, g_t^{\alpha \beta} \, \delta_O\Pi(\omega,\omega^\prime),
\nonumber
\end{eqnarray}
where $O=\mathcal{K}$ or $\mathcal{S}$. At the hadronic level, these correlation functions contain the double-pole contributions
\begin{align}
\delta_{\mathcal K}\Pi(\omega,\omega^\prime)&=\frac{f^2K}{(\overline{\Lambda}-\omega)(\overline{\Lambda}-\omega^\prime)}+\cdots,
\nonumber\\
\delta_{\mathcal S}\Pi(\omega,\omega^\prime)&=\frac{d_Mf^2\Sigma}{(\overline{\Lambda}-\omega)(\overline{\Lambda}-\omega^\prime)}+\cdots.
\label{eq:KS-correlators}
\end{align}
After evaluating these correlation functions using the OPE and performing the double Borel transformation, the corresponding sum rules can be used to determine $K$ and $\Sigma$. For the $[\Omega_b(3/2^-),2,1,\lambda]$ baryon, they are explicitly given by
\begin{eqnarray}
&&f^2 K\, e^{-\overline{\Lambda} / T}
\nonumber\\&=& \int_{2m_s}^{\omega_c} e^{-\omega/T} \left(-\frac{2\omega^9}{45\pi^4}+\frac{74 m_s^2 \omega^7}{105\pi^4}\right) d\omega \nonumber\\&&
 -\frac{352\,m_s\langle\bar{s}s\rangle\,T^6}{3\pi^2}+\frac{205\langle g^2GG\rangle\,T^6}{72\pi^4} \nonumber\\&&
-\frac{37\langle g^2GG\rangle\,m_s^2 T^4}{72\pi^4}-\frac{13\langle g_s \bar{s} \sigma G s\rangle^2}{72} \nonumber\\&&
 +\frac{7\,m_s\langle\bar{s}s\rangle\langle g^2GG\rangle\,T^2}{432\pi^2}-\frac{5\langle g^2GG\rangle\langle\bar{s}s\rangle\langle g_s \bar{s} \sigma G s\rangle}{6912\,T^2} \nonumber\\&&
 -\frac{\langle g^2GG\rangle\,m_s^2\langle\bar{s}s\rangle^2}{10368\,T^2}-\frac{5\langle g^2GG\rangle\langle g_s \bar{s} \sigma G s\rangle^2}{110592\,T^4}\,.
\label{eq:Kc}
\\
&&f^2 d_M \Sigma\, e^{-\overline{\Lambda} / T}
\nonumber\\&=&\frac{5\langle g^2GG\rangle\,T^6}{3\pi^4}
-\frac{5\langle g^2GG\rangle\,m_s^2 T^4}{16\pi^4}\nonumber\\&&
+\frac{5\,m_s\langle\bar{s}s\rangle\langle g^2GG\rangle\,T^2}{144\pi^2}\,.
\label{eq:Sc}
\end{eqnarray}

Combining the leading-order contribution and the ${\mathcal O}(1/m_b)$ corrections, the mass of the $\Omega_b(3/2^-)$ baryon is obtained as
\begin{equation}
m_{\Omega_b(3/2^-)}
=
m_b
+\overline{\Lambda}
-\frac{1}{2m_b}
\left(
K
+d_{3/2,2}C_{\rm mag}\Sigma
\right).
\label{eq:heavy-baryon-mass}
\end{equation}
For the two members of the $[\Omega_b,2,1,\lambda]$ doublet, the kinetic-energy contribution is the same, whereas the chromomagnetic interaction lifts their degeneracy and generates the intra-doublet mass splitting,
\begin{equation}
\Delta m
=
m_{\Omega_b(5/2^-)}
-
m_{\Omega_b(3/2^-)}
=
\frac{5}{m_b}C_{\rm mag}\Sigma .
\label{eq:doublet-splitting}
\end{equation}
The explicit $1/m_b$ dependence naturally leads to smaller intra-doublet mass splittings in the bottom sector than in the corresponding charm sector.

We perform the QCD sum rules analyses for all four $P$-wave flavor-sextet HQET multiplets introduced in Sec.~\ref{sec:classification}. The resulting masses and intra-doublet mass splittings of the charmed and bottom baryons are summarized in Table~\ref{tab:mass}. The calculated mass spectrum provides an independent theoretical test of the phenomenological assignments proposed in Sec.~\ref{sec:classification}. In particular, the intra-doublet mass splittings provide useful constraints because they are generally more stable than the absolute masses against variations of the heavy-quark mass and other systematic uncertainties. The substantially smaller splittings obtained in the bottom sector further suggest that two nearby members of the same HQET doublet may remain experimentally unresolved.

\section{Strong decay properties}
\label{sec:decay}

In this section, we investigate the strong decays of the $P$-wave flavor-sextet bottom baryons using light-cone sum rules within heavy quark effective theory (HQET), following the framework developed in Refs.~\cite{Chen:2017sci,Yang:2019cvw}. The updated mass parameters obtained in Sec.~\ref{sec:mass} are used as inputs in the decay analyses. 

As in Sec.~\ref{sec:mass}, we take the $\Omega_b(3/2^-)$ state belonging to the $[{\mathbf 6}_F,2,1,\lambda]$ multiplet as a representative example and consider the charged $D$-wave decay channel
\begin{equation}
\Omega_b^-(3/2^-)\to\Xi_b^0(1/2^+)+K^-.
\label{eq:example-decay}
\end{equation}
The corresponding coupling constant $g_D$ is defined through the effective Lagrangian
\begin{equation}
\mathcal{L}^{D}_{\Omega_b^-(3/2^-)\to\Xi_b^0K^-}
=
g_D
\times
\bar\Omega^-_{b\mu}\gamma_\nu\gamma_5\Xi_b^0
\,\partial^\mu\partial^\nu K^- .
\label{eq:example-lagrangian}
\end{equation}

To determine the coupling constant $g_D$, we consider the light-cone correlation function
\begin{eqnarray}
&&\Pi^\alpha(\omega,\omega^\prime)
\nonumber\\
&=&
i\int d^4x\,e^{-ik\cdot x}
\langle0|T[J^\alpha_{\Omega^-_b(3/2^-)}(0)\bar J_{\Xi_b^0}(x)]|K^-(q)\rangle
\nonumber\\
&=&
\frac{1+v\!\!\!\slash}{2}
\times
G^\alpha(\omega,\omega^\prime),
\label{eq:decay-correlator}
\end{eqnarray}
where $k^\prime=k+q$, $\omega=v\cdot k$, and $\omega^\prime=v\cdot k^\prime$. The interpolating current for the initial $\Omega_b^-(3/2^-)$ baryon has been given in Eq.~(\ref{eq:current-omegab32}), while the ground-state $\Xi_b^0(1/2^+)$ baryon is interpolated by
\begin{equation}
J_{\Xi_b^0}
=
\epsilon_{abc}\left[u^{aT}C\gamma_5s^b\right]h_v^c .
\label{eq:decay-current-xib}
\end{equation}
At the hadronic level, the contribution from the lowest-lying $\Omega_b^-(3/2^-)$ and $\Xi_b^0(1/2^+)$ states can be written as
\begin{eqnarray}
G^\alpha(\omega,\omega^\prime)
&=&
g_D
\frac{f_{\Omega^-_b(3/2^-)}f_{\Xi_b^0}}
{(\overline{\Lambda}_{\Omega^-_b(3/2^-)}-\omega^\prime)(\overline{\Lambda}_{\Xi_b^0}-\omega)}
\times q^\alpha \, q\!\!\!\slash \,\gamma_5
\nonumber\\
&&~~~~~~~~~~~~~~~~~~~~~~~~~~~~~~~~~+\cdots ,
\label{eq:decay-hadron}
\end{eqnarray}
where the ellipsis denotes contributions from higher resonances and continuum states.

At the quark-gluon level, the same correlation function is evaluated near the light cone in terms of the kaon light-cone distribution amplitudes. After performing the double Borel transformation, $\omega^\prime\to T_1$ and $\omega\to T_2$, and subtracting the continuum contributions, we obtain the light-cone sum rule for the coupling constant $g_D$:
\begin{eqnarray}
&& g_D f_{\Omega_b^-({3/2}^-)} f_{\Xi_b^{0}} e^{- {\bar \Lambda_{\Omega_b^-({3/2}^-)} \over T_1}} e^{ - {\bar \Lambda_{\Xi_b^{0}} \over T_2}}
\label{eq:621lambda}
\\ \nonumber &=& 8 \times \Big ( -\frac{i f_K m_s u_0}{4\pi^2}T^3 f_2({\omega_c \over T})\phi_{2;K}(u_0)
\\ \nonumber &&-\frac{i f_K m_K^2 u_0}{12(m_u+m_s)\pi^2}T^3 f_2({\omega_c \over T})\phi_{3;K}^\sigma(u_0)
\\ \nonumber &&+\frac{i f_K m_s u_0}{64\pi^2}T f_0({\omega_c \over T})\phi_{4;K}(u_0)
\\ \nonumber &&+\frac{i f_K u_0}{12}\langle \bar s s\rangle T f_0({\omega_c \over T})\phi_{2;K}(u_0)
\\ \nonumber &&-\frac{i f_K m_s u_0}{288(m_u+m_s)}\langle \bar s s\rangle {1\over T}\phi_{3;K}^\sigma(u_0)
\\ \nonumber &&-\frac{i f_K u_0}{192}\langle \bar s s\rangle {1\over T}\phi_{4;K}(u_0)
\\ \nonumber &&
-\frac{i f_K u_0}{192}\langle g_s \bar s\sigma G s\rangle {1\over T}\phi_{2;K}(u_0)
\\ \nonumber &&+\frac{i f_K u_0}{3072}\langle g_s \bar s \sigma G s\rangle{1\over T^3}\phi_{4;K}(u_0) \Big )
\\ \nonumber &-&
\Big(-\frac{i f_{3K}}{2\pi^2}T^3f_2({\omega_c \over T}) \int_0^{1 \over 2} d\alpha_2 \int_{{1 \over 2}-\alpha_2}^{1-\alpha_2} d\alpha_3 ({u_0 \over \alpha_3} \Phi_{3;K}(\underline{\alpha})
\\ \nonumber &&-{1 \over \alpha_3} \Phi_{3;K}(\underline{\alpha}))
+\frac{i f_{3K}}{2\pi^2}T^3f_2({\omega_c\over T}) \times
\\ \nonumber &&\int_0^{1 \over 2} d\alpha_2 \int_{{1 \over 2}-\alpha_2}^{1-\alpha_2} d\alpha_3 {1\over \alpha_3}{\partial\over\partial\alpha_3}(\alpha_3 u_0\Phi_{3;K}(\underline{\alpha})
\\ \nonumber &&+\alpha_2\Phi_{3;K}(\underline{\alpha})-\Phi_{3;K}(\underline{\alpha}))\Big ) \, .
\end{eqnarray}
We work at the symmetric point $T_1=T_2=2T$, for which $u_0=\frac{T_2}{T_1+T_2}=\frac{1}{2}$. The continuum-subtraction functions are defined as $f_n(x)\equiv1-e^{-x}\sum_{k=0}^{n}\frac{x^k}{k!}$. The explicit forms of the kaon light-cone distribution amplitudes entering the sum rule are taken from Refs.~\cite{Ball:1998je,Ball:2006wn,Ball:2004rg,Ball:1998kk,Ball:1998sk,Ball:1998ff,Ball:2007rt,Ball:2007zt}.

For the representative channel considered above, we obtain
\begin{equation}
g_D
=
7.37^{+3.83}_{-2.89}~{\rm GeV}^{-2}.
\label{eq:example-coupling}
\end{equation}
The corresponding decay amplitude is
\begin{equation}
\mathcal{M}\left[\Omega_b^-(3/2^-)\to\Xi_b^0K^-\right]
=
g_D \, q_\mu
\times
\bar u_{\Xi_b^0} q\!\!\!\slash \gamma_5u_{\Omega_b^-}^{\mu} .
\label{eq:example-amplitude}
\end{equation}
Including the isospin-related channel, the total $\Xi_bK$ decay width is
\begin{eqnarray}
\Gamma[\Omega_b(3/2^-)\to\Xi_bK]
&=&
2\times\Gamma[\Omega_b^-(3/2^-)\to\Xi_b^0K^-]
\nonumber\\
&=&
1.8^{+2.3}_{-1.1}~{\rm MeV}.
\label{eq:example-width-result}
\end{eqnarray}
This result is obtained for the $\Omega_b(3/2^-)$ state belonging to the $[\mathbf{6}_F,2,1,\lambda]$ multiplet.

The same procedure is applied to the other kinematically allowed strong decay channels. The resulting partial and total decay widths of the $P$-wave flavor-sextet bottom baryons are summarized in Table~\ref{tab:decaybottom}. The predicted widths exhibit a clear dependence on the HQET configuration and total spin. In particular, the $[\mathbf{6}_F,2,1,\lambda]$ states can have sizable $D$-wave transitions into ground-state antitriplet bottom baryons, whereas the corresponding $\Omega_b$ states remain narrow because of the limited phase space. These decay patterns provide information complementary to the mass spectrum obtained in Sec.~\ref{sec:mass} and will be used together with possible configuration mixing in Sec.~\ref{sec:mixing} to interpret the observed bottom baryons.

\section{Configuration mixing}
\label{sec:mixing}

In the heavy quark limit, the angular momentum $j_l$ of the light degrees of freedom is conserved. At finite heavy-quark mass, however, $j_l$ is no longer an exact quantum number, and states with the same flavor and $J^P$ can mix through ${\mathcal O}(1/m_Q)$ effects. More generally, configuration mixing may also arise from the breaking of approximate flavor symmetries. An example is the flavor-$SU(3)$-breaking mixing between the predominantly $\overline{\mathbf{3}}_F$ $\Xi_Q$ and $\mathbf{6}_F$ $\Xi_Q^\prime$ states~\cite{Chen:2025way}. The HQET assignments discussed above should therefore be understood as specifying the dominant components of the physical states.

Since mixing among different HQET multiplets is induced by finite-heavy-quark-mass effects, it is expected to be considerably smaller in the bottom sector than in the charm sector. We therefore discuss possible mixing in the bottom sector only qualitatively, while explicitly implementing the mixing between the $J^P=3/2^-$ members of the $[\mathbf{6}_F,1,1,\lambda]$ and $[\mathbf{6}_F,2,1,\lambda]$ multiplets in the charm sector.

\subsection{Bottom baryons}
\label{subsec:mixing-bottom}

For the bottom baryons, small configuration mixing is particularly relevant when it can activate decay channels that vanish or are strongly suppressed in the pure-HQET limit. This is especially important in the $\Omega_b$ sector. As shown in Table~\ref{tab:decaybottom}, among the six narrow $P$-wave $\Omega_b$ states considered here, only the $J^P=3/2^-$ and $5/2^-$ members of the $[\mathbf{6}_F,2,1,\lambda]$ multiplet have appreciable $\Xi_bK$ widths, whereas the $[\mathbf{6}_F,1,1,\lambda]$ and $[\mathbf{6}_F,1,0,\rho]$ states have strongly suppressed decays into this channel.

Our light-cone sum rules calculation finds a large $S$-wave $\Xi_bK$ decay width for the broad $[\Omega_b(1/2^-),0,1,\lambda]$ baryon:
\begin{equation}
\Gamma[[\Omega_b(1/2^-),0,1,\lambda]\to\Xi_bK]
=1400~{\rm MeV},
\end{equation}
Consequently, even a small admixture of this broad configuration may induce observable $\Xi_bK$ decay amplitudes for the two narrow $\Omega_b(1/2^-)$ baryons belonging to the $[\mathbf{6}_F,1,1,\lambda]$ and $[\mathbf{6}_F,1,0,\rho]$ multiplets. Such small admixtures can modify the experimentally accessible decay modes without substantially changing the dominant HQET components.

\subsection{Charmed baryons}
\label{subsec:mixing-charm}

Finite-heavy-quark-mass effects are more pronounced in the charm sector~\cite{Yang:2020zjl}. Of particular importance is the mixing between the two $J^P=3/2^-$ states belonging to the $[\mathbf{6}_F,1,1,\lambda]$ and $[\mathbf{6}_F,2,1,\lambda]$ multiplets. For each flavor sector, $B_c=\Sigma_c,\Xi_c^\prime,\Omega_c$, we define the two physical $J^P=3/2^-$ states as
\begin{align}
&\begin{pmatrix}
|B_c(3/2^-)\rangle_L\\
|B_c(3/2^-)\rangle_H
\end{pmatrix}
\nonumber \\&~~~~~=
\begin{pmatrix}
\cos\theta & \sin\theta\\
-\sin\theta & \cos\theta
\end{pmatrix}
\begin{pmatrix}
|[B_c(3/2^-),1,1,\lambda]\rangle\\
|[B_c(3/2^-),2,1,\lambda]\rangle
\end{pmatrix},
\label{eq:mixing}
\end{align}
where the subscripts $L$ and $H$ denote the lower- and higher-mass physical states, respectively.

We determine the mixing angle phenomenologically by requiring a simultaneous description of the masses and strong decay properties of $\Xi_c(2923)$, $\Xi_c(2939)$, $\Omega_c(3050)$, and $\Omega_c(3066)$. Combining the QCD sum rules mass spectrum with the strong decay analyses, we find a preferred mixing angle
\begin{equation}
\theta=37^\circ\pm5^\circ.
\label{eq:common-angle}
\end{equation}
For this mixing angle, the lower-mass states $\Xi_c(2923)$ and $\Omega_c(3050)$ remain dominated by the $[\mathbf{6}_F,1,1,\lambda]$ configuration, whereas the higher-mass $\Xi_c(2939)$ and $\Omega_c(3066)$ states remain dominated by the $[\mathbf{6}_F,2,1,\lambda]$ configuration. 

We then apply the same mixing pattern to the corresponding $\Sigma_c(3/2^-)$ states. The resulting masses, decay properties, and possible experimental assignments are summarized in Table~\ref{tab:decaycharm}. This mixing allows each physical $J^P=3/2^-$ state to inherit decay amplitudes from both HQET basis configurations, thereby improving the simultaneous description of the observed masses and widths. Small admixtures of other configurations with the same flavor and $J^P$, parameterized by the small mixing angles $\theta^\prime$ and $\theta^{\prime\prime}$ in Table~\ref{tab:decaycharm}, are also allowed at finite charm-quark mass. Although they have little impact on the masses and total decay widths, they can generate otherwise vanishing or strongly suppressed decay amplitudes, thereby allowing the dominant HQET assignments to remain compatible with the experimentally observed decay channels.

\section{Summary and discussion}
\label{sec:summary}

In this work, we have systematically investigated the $P$-wave heavy baryons in the $SU(3)$ flavor $\mathbf{6}_F$ representation. Within heavy quark effective theory (HQET), the seven states in each of the $\Sigma_Q$, $\Xi_Q^\prime$, and $\Omega_Q$ sectors ($Q=c,b$) are classified into four multiplets, $[\mathbf{6}_F,0,1,\lambda]$, $[\mathbf{6}_F,1,1,\lambda]$, $[\mathbf{6}_F,2,1,\lambda]$, and $[\mathbf{6}_F,1,0,\rho]$. We have studied their masses using QCD sum rules, their strong decay properties using light-cone sum rules, and configuration mixing among states with the same flavor and $J^P$. Combining these results with the experimentally observed mass patterns, we obtain a unified description of the $P$-wave flavor-sextet heavy-baryon spectrum. We note that the results for the charm sector have been presented in our recent study~\cite{Luo:2026elv}. These results are summarized here for completeness and for a unified comparison with the bottom sector.

In the charm sector, a particularly striking feature is the similarity between the mass patterns of the four excited $\Xi_c$ structures, $\Xi_c(2882)$, $\Xi_c(2923)$, $\Xi_c(2939)$, and $\Xi_c(2965)$, and the four lower narrow $\Omega_c$ structures, $\Omega_c(3000)$, $\Omega_c(3050)$, $\Omega_c(3066)$, and $\Omega_c(3090)$. Their successive experimental mass splittings are approximately
\begin{align}
\Delta M_{\Xi_c}
&\simeq (41,\,16,\,26)~{\rm MeV},
\nonumber\\
\Delta M_{\Omega_c}
&\simeq (50,\,16,\,24)~{\rm MeV}.
\end{align}
The correspondence between these two sequences strongly suggests a common underlying HQET structure. We interpret them predominantly as the four $\lambda$-mode states belonging to the $[\mathbf{6}_F,1,1,\lambda]$ and $[\mathbf{6}_F,2,1,\lambda]$ multiplets. In particular, the two $J^P=3/2^-$ configurations can mix at finite charm-quark mass. A simultaneous description of the masses and strong decay properties of $\Xi_c(2923)$, $\Xi_c(2939)$, $\Omega_c(3050)$, and $\Omega_c(3066)$ favors a common mixing angle $\theta=37^\circ\pm5^\circ$. For this preferred mixing, the lower-mass states remain dominated by the $[\mathbf{6}_F,1,1,\lambda]$ configuration, whereas the higher-mass states remain dominated by the $[\mathbf{6}_F,2,1,\lambda]$ configuration.

The additional narrow $\Omega_c(3119)$ is tentatively associated with the $J^P=3/2^-$ member of the $[\mathbf{6}_F,1,0,\rho]$ doublet. Its $J^P=1/2^-$ partner, as well as the $[\Omega_c(1/2^-),0,1,\lambda]$ state, can be considerably broader and therefore more difficult to identify experimentally. The situation in the $\Sigma_c$ sector is less transparent, with several nearby $P$-wave states potentially having sizable widths and contributing to the observed $\Sigma_c(2800)$ and $\Sigma_c(2900)$ structures.

In the bottom sector, the heavier bottom-quark mass leads to substantially smaller spin-dependent splittings. The $\Sigma_b(6097)$ has been observed in the $\Lambda_b\pi$ channel, while the $\Xi_b(6227)$ has been observed in both the $\Lambda_bK$ and $\Xi_b\pi$ channels. These decay patterns are consistent with the $[\mathbf{6}_F,2,1,\lambda]$ assignments, for which both the $J^P=3/2^-$ and $5/2^-$ members can decay into the corresponding experimentally observed channels. Moreover, the QCD sum rules predict intra-doublet mass splittings of only about $10$--$12$~MeV, comparable to or smaller than the calculated strong decay widths. The observed $\Sigma_b(6097)$ and $\Xi_b(6227)$ structures may therefore each contain unresolved contributions from both members of the corresponding doublet. The four $\Sigma_b$ and $\Xi_b^\prime$ members of the lower $[\mathbf{6}_F,1,1,\lambda]$ doublet have not yet been identified experimentally. Their predicted dominant decay modes, $\Sigma_b\pi$, $\Sigma_b^*\pi$, $\Xi_b^\prime\pi$, and $\Xi_b^*\pi$, provide promising channels for future searches.

The $\Omega_b$ sector exhibits an especially rich pattern. The six narrow $P$-wave states belonging to the $[\mathbf{6}_F,1,1,\lambda]$, $[\mathbf{6}_F,2,1,\lambda]$, and $[\mathbf{6}_F,1,0,\rho]$ multiplets provide a natural framework for interpreting the four narrow structures observed experimentally. We tentatively associate $\Omega_b(6316)$ predominantly with the $[\Omega_b(1/2^-),1,1,\lambda]$ state, while its $J^P=3/2^-$ partner is expected to lie nearby but has not yet been identified experimentally. The $\Omega_b(6330)$ and $\Omega_b(6340)$ are predominantly associated with the $J^P=3/2^-$ and $5/2^-$ members of the $[\mathbf{6}_F,2,1,\lambda]$ doublet, respectively. The two $\Omega_b$ members of the $[\mathbf{6}_F,1,0,\rho]$ doublet are both predicted to be narrow, with a mass splitting of only about $2$~MeV, and may remain experimentally unresolved within the $\Omega_b(6350)$ structure. Configuration mixing may play an important role in the experimental visibility of the narrow $\Omega_b$ states. In particular, the broad $[\Omega_b(1/2^-),0,1,\lambda]$ configuration has a large $S$-wave $\Xi_bK$ decay width. Even a small admixture of this configuration can therefore induce nonvanishing $\Xi_bK$ decay amplitudes for the two narrow $[\Omega_b(1/2^-),1,1,\lambda]$ and $[\Omega_b(1/2^-),1,0,\rho]$ states, without substantially changing their dominant HQET components.

The main experimental consequences and predictions of the above picture, as illustrated in Fig.~\ref{fig:classification}, can be summarized as follows:
\begin{itemize}

\item Four relatively narrow $P$-wave $\Sigma_c$ states are expected in the relevant mass region. The observed $\Sigma_c(2800)$ and $\Sigma_c(2900)$ structures may therefore contain several overlapping resonances and could eventually be resolved into up to four distinct resonances.

\item Four relatively narrow $P$-wave $\Xi_c^\prime$ states are expected and can naturally account for the observed $\Xi_c(2882)$, $\Xi_c(2923)$, $\Xi_c(2939)$, and $\Xi_c(2965)$ structures.

\item Five relatively narrow $P$-wave $\Omega_c$ states are expected and can naturally account for the observed $\Omega_c(3000)$, $\Omega_c(3050)$, $\Omega_c(3066)$, $\Omega_c(3090)$, and $\Omega_c(3119)$ structures.

\item Four relatively narrow $P$-wave $\Sigma_b$ states are expected. The $\Sigma_b(6097)$ structure may eventually be resolved into two nearby states corresponding predominantly to the $J^P=3/2^-$ and $5/2^-$ members of the $[\mathbf{6}_F,2,1,\lambda]$ doublet, with masses around $6097$ and $6109$~MeV, respectively. The two lower $[\mathbf{6}_F,1,1,\lambda]$ states may be searched for in the $\Sigma_b\pi$ and $\Sigma_b^*\pi$ channels.

\item Four relatively narrow $P$-wave $\Xi_b^\prime$ states are expected. The $\Xi_b(6227)$ structure may eventually be resolved into two nearby states corresponding predominantly to the $J^P=3/2^-$ and $5/2^-$ members of the $[\mathbf{6}_F,2,1,\lambda]$ doublet, with masses around $6225$ and $6236$~MeV, respectively. The two lower $[\mathbf{6}_F,1,1,\lambda]$ states may be searched for in the $\Xi_b^\prime\pi$ and $\Xi_b^*\pi$ channels.

\item Six narrow $P$-wave $\Omega_b$ states are expected. The $\Omega_b(6316)$ is predominantly associated with the $J^P=1/2^-$ member of the $[\mathbf{6}_F,1,1,\lambda]$ doublet, while its $J^P=3/2^-$ partner has not yet been identified experimentally. The $\Omega_b(6330)$ and $\Omega_b(6340)$ are predominantly associated with the $J^P=3/2^-$ and $5/2^-$ members of the $[\mathbf{6}_F,2,1,\lambda]$ doublet, respectively. The $\Omega_b(6350)$ may contain the two unresolved members of the $[\mathbf{6}_F,1,0,\rho]$ doublet, whose mass splitting is predicted to be only about $2$~MeV.

\item Motivated by the mass pattern of the corresponding four $\lambda$-mode states in the charm sector, the missing $[\Omega_b(3/2^-),1,1,\lambda]$ state may lie around $6327$~MeV. Together with the observed $\Omega_b(6316)$, $\Omega_b(6330)$, $\Omega_b(6340)$, and $\Omega_b(6350)$ structures, this would lead to the successive mass splittings:
\begin{align}
\Delta M_{\Omega_b}
&\simeq (11,\,3,\,10,\,10)~{\rm MeV},
\end{align}
which roughly resemble a scaled-down version of the mass-splitting pattern among the five observed $\Omega_c$ states:
\begin{align}
\Delta M_{\Omega_c}
&\simeq (50,\,16,\,24,\,29)~{\rm MeV}.
\end{align}

\end{itemize}
Detailed predictions for the masses and decay properties of the observed and missing states are summarized in Tables~\ref{tab:decaycharm} and \ref{tab:decaybottom}.

The combined charm and bottom analyses show that the experimentally observed heavy-baryon spectrum need not exhibit a one-to-one correspondence with the underlying HQET multiplets. Finite-heavy-quark-mass effects generate mass splittings and configuration mixing, while strong decay dynamics determine whether a state appears as an isolated narrow peak, overlaps with nearby states, or becomes too broad to be readily identified. The observed spectrum can therefore be viewed as a dynamically filtered manifestation of the underlying HQET structure. More precise measurements of the masses, line shapes, spin-parity quantum numbers, and decay modes of the excited heavy baryons, together with dedicated searches for the missing states, will be important for testing this spectroscopic picture and clarifying how the HQET multiplet structure is realized in the physical spectrum.

\begin{table*}[hbt]
\begin{center}
\renewcommand{\arraystretch}{1.5}
\caption{Strong decay properties of the $P$-wave flavor-sextet bottom baryons, with configuration mixing not explicitly included in the quoted decay widths. The $\Sigma_b$ and $\Xi_b^\prime$ states belonging to the $[\mathbf{6}_F,0,1,\lambda]$ and $[\mathbf{6}_F,1,0,\rho]$ multiplets are predicted to be broad and are therefore omitted for compactness, while only non-negligible partial decay widths are quoted numerically. $\Gamma_S$ and $\Gamma_D$ denote the $S$- and $D$-wave partial widths, respectively. A small admixture of the broad $[\Omega_b(1/2^-),0,1,\lambda]$ configuration, which has a large $S$-wave $\Xi_bK$ decay width, can induce nonvanishing $\Xi_bK$ decay amplitudes for the narrow $\Omega_b(1/2^-)$ states in the $[\mathbf{6}_F,1,1,\lambda]$ and $[\mathbf{6}_F,1,0,\rho]$ multiplets. These mixing-induced $\Xi_bK$ channels are marked by $\neq0$.}
\begin{tabular}{ c | c | c | c | c | c | c }
\hline\hline
\multirow{2}{*}{~Multiplet~} & ~Baryon~ & ~~~Mass~~~ & Splitting & \multirow{2}{*}{~~~~~~~~~~~Decay channel~~~~~~~~~~~ }& Total width  & \multirow{2}{*}{~Candidate~}
\\& ($J^P$) & ({GeV}) & ({MeV}) & & ({MeV}) &
\\ \hline\hline
\multirow{4}{*}{$[\mathbf{6}_F, 1, 1, \lambda]$} & $\Sigma_b({1\over2}^-)$ & $6.075\pm0.128$& \multirow{4}{*}{$6\pm3$} &
$\begin{array}{c}
\Gamma_S\left(\Sigma_b({1\over2}^-)\to \Sigma_b\pi\right)=15.3~{^{+23.8}_{-15.3}}~{\rm MeV} \\
\Gamma_D\left(\Sigma_b({1\over2}^-)\to \Sigma_b^{*}\pi\right)=0.12~{^{+1.80}_{-0.12}}~{\rm MeV}
\end{array}$ &$15.4~{^{+25.6}_{-15.4}}$&--
\\ \cline{2-3}\cline{5-7}
                                                                           & $\Sigma_b({3\over2}^-)$ & $6.081 \pm0.129$&&
$\begin{array}{c}
\Gamma_D\left(\Sigma_b({3\over2}^-)\to \Sigma_b\pi\right)=0.72~{^{+7.01}_{-0.72}}~{\rm MeV} \\
\Gamma_S\left(\Sigma_b({3\over2}^-)\to \Sigma_b^{*}\pi\right)=4.2~{^{+6.8}_{-4.2}}~{\rm MeV} \\
\Gamma_D\left(\Sigma_b({3\over2}^-)\to \Sigma_b^{*}\pi\right)=0.10~{^{+1.27}_{-0.10}}~{\rm MeV} 
\end{array}$  &$5.0~{^{+15.1}_{-5.0}}$&--
\\ \cline{1-7}
\multirow{4}{*}{$[\mathbf{6}_F,2,1,\lambda]$}&$\Sigma_b({3\over2}^-)$&$6.089\pm 0.182$&\multirow{4}{*}{$12\pm6$}&
$\begin{array}{c}
\Gamma_D\left(\Sigma_b({3\over2}^-)\to\Lambda_b\pi\right)=51.1~{^{+73.4}_{-34.4}}~{\rm MeV}\\
\Gamma_D\left(\Sigma_b({3\over2}^-)\to \Sigma_b \pi\right)=1.7~{^{+2.4}_{-1.2}}~{\rm MeV}\\
\Gamma_S\left(\Sigma_b({3\over2}^-)\to \Sigma_b^{*}\pi\right)=0.03~{^{+0.08}_{-0.03}}~{\rm MeV}\\
\Gamma_D\left(\Sigma_b({3\over2}^-)\to \Sigma_b^{*}\pi\right)=0.24~{^{+0.34}_{-0.17}}~{\rm MeV}
\end{array}$
&$53.1~{^{+76.2}_{-35.8}}$& \multirow{5}{*}{$\Sigma_b(6097)$}
\\ \cline{2-3} \cline{5-6}
&$\Sigma_b({5\over2}^-)$&$6.101 \pm0.180 $&&$\begin{array}{c}
\Gamma_D\left(\Sigma_b({5\over2}^-)\to\Lambda_b\pi\right)=7.7~{^{+11.1}_{-5.2}}~{\rm MeV}\\
\Gamma_D\left(\Sigma_b({5\over2}^-)\to \Sigma_b \pi\right)=0.06~{^{+0.08}_{-0.04}}~{\rm MeV}\\
\Gamma_D\left(\Sigma_b({5\over2}^-)\to \Sigma_b^{*}\pi\right)=1.3~{^{+1.9}_{-0.9}}~{\rm MeV}
\end{array}$&$9.1~{^{+13.1}_{-6.1}}$&
\\ \hline \hline
\multirow{5}{*}{$[\mathbf{6}_F,1,1,\lambda]$}&$\Xi^\prime_b({1\over2}^-)$&$6.206\pm0.111 $ &\multirow{5}{*}{$7 \pm 2$}&
$\begin{array}{c}
\Gamma_S\left(\Xi_b^{\prime}({1\over2}^-)\to \Xi_b^{\prime}\pi\right)=4.4~{^{+6.1}_{-4.3}}~{\rm MeV}\\
\Gamma_D\left(\Xi_b^{\prime}({1\over2}^-)\to\Xi_b^{*}\pi\right)=0.14~{^{+1.27}_{-0.14}}~{\rm MeV}
\end{array}$&$4.5~{^{+7.4}_{-4.4}}$&--
\\ \cline{2-3}\cline{5-7}
&$\Xi_b^{\prime}({3\over2}^-)$&$6.213 \pm 0.111$&&
$\begin{array}{c}
\Gamma_D\left(\Xi_b^{\prime}({3\over2}^-)\to\Xi_b^{\prime}\pi\right)=0.29~{^{+1.89}_{-0.29}}~{\rm MeV}\\
\Gamma_S\left(\Xi_b^{\prime}({3\over2}^-)\to\Xi_b^{*}\pi\right)=1.3~{^{+1.7}_{-1.3}}~{\rm MeV}\\
\Gamma_D\left(\Xi_b^{\prime}({3\over2}^-)\to\Xi_b^{*}\pi\right)=0.04~{^{+0.03}_{-0.04}}~{\rm MeV}
\end{array}$&$1.6~{^{+3.6}_{-1.6}}$&--
\\ \cline{1-7}
\multirow{6}{*}{$[\mathbf{6}_F,2,1,\lambda]$}&$\Xi_b^{\prime}({3\over2}^-)$&$6.218 \pm0.157$&\multirow{3}{*}{$11\pm5$}&
$\begin{array}{c}
\Gamma_D\left(\Xi_b^{\prime}({3\over2}^-)\to \Xi_b\pi\right)=19.0~{^{+24.9}_{-12.5}}~{\rm MeV}\\
\Gamma_D\left(\Xi_b^{\prime}({3\over2}^-)\to \Lambda_b K\right)=7.3~{^{+10.4}_{-4.8}}~{\rm MeV}\\
\Gamma_D\left(\Xi_b^{\prime}({3\over2}^-)\to \Xi_b^{\prime}\pi\right)=0.78~{^{+1.03}_{-0.52}}~{\rm MeV}\\
\Gamma_S\left(\Xi_b^{\prime}({3\over2}^-)\to \Xi_b^{*}\pi\right)=0.01~{^{+0.02}_{-0.01}}~{\rm MeV}\\
\Gamma_D\left(\Xi_b^{\prime}({3\over2}^-)\to \Xi_b^{*}\pi\right)=0.12~{^{+0.16}_{-0.08}}~{\rm MeV}
\end{array}$&$27.2~{^{+36.5}_{-17.9}}$& \multirow{5}{*}{$\Xi_b(6227)$}
\\ \cline{2-3} \cline{5-6}
&$\Xi^\prime_b({5\over2}^-)$&$6.230 \pm0.155$&&
$\begin{array}{c}
\Gamma_D\left(\Xi_b^{\prime}({5\over2}^-)\to \Xi_b\pi\right)=1.3~{^{+1.7}_{-0.8}}~{\rm MeV}\\
\Gamma_D\left(\Xi_b^{\prime}({5\over2}^-)\to \Lambda_b K\right)=1.2~{^{+1.7}_{-0.8}}~{\rm MeV}\\
\Gamma_D\left(\Xi_b^{\prime}({5\over2}^-)\to \Xi_b^{\prime}\pi\right)=0.03~{^{+0.07}_{-0.03}}~{\rm MeV}\\
\Gamma_D\left(\Xi_b^{\prime}({5\over2}^-)\to \Xi_b^{*}\pi\right)=0.64~{^{+2.49}_{-0.64}}~{\rm MeV}
\end{array}$
&$3.2~{^{+6.0}_{-2.3}}$&
\\ \hline \hline
$[\mathbf{6}_F,0,1,\lambda]$&$\Omega_b({1\over2}^-)$&$6.302 \pm 0.119$ &--&
$\Gamma_S\left(\Omega_b({1\over2}^-)\to \Xi_b K\right)=1400~{\rm MeV}$
&$1400$&--
\\\hline
\multirow{2}{*}{$[\mathbf{6}_F,1,1,\lambda]$}&$\Omega_b({1\over2}^-)$&$6.315 \pm 0.092$&\multirow{2}{*}{$7 \pm 2$}&
$\Gamma_S\left(\Omega_b({1\over2}^-)\to \Xi_b K\right) \neq 0$
&$\sim~0$&$\Omega_b(6316)$
\\ \cline{2-3} \cline{5-7}
&$\Omega_b({3\over2}^-)$&$6.322 \pm 0.091$&&--&$\sim~0$&--
\\ \cline{1-7}
\multirow{2}{*}{$[\mathbf{6}_F,2,1,\lambda]$}&$\Omega_b({3\over2}^-)$& $6.331\pm0.126 $&\multirow{2}{*}{$10\pm5$}&
$\begin{array}{c}
\Gamma_D\left(\Omega_b({3\over2}^-)\to \Xi_b K\right)=1.8~{^{+2.3}_{-1.1}}~{\rm MeV}
\end{array}$&$1.8~{^{+2.3}_{-1.1}}$&$\Omega_b(6330)$
\\ \cline{2-3} \cline{5-7}
&$\Omega_b({5\over2}^-)$&$6.341 \pm0.125 $&&$\begin{array}{c}
\Gamma_D\left(\Omega_b({5\over2}^-)\to \Xi_b K\right)=0.39~{^{+0.51}_{-0.25}}~{\rm MeV}
\end{array}$ &$0.39~{^{+0.51}_{-0.25}}$&$\Omega_b(6340)$
\\ \cline{1-7}
\multirow{2}{*}{$[\mathbf{6}_F,1,0,\rho]$}&$\Omega_b({1\over2}^-)$&$6.351 \pm0.128 $&\multirow{2}{*}{$2 \pm 1$}&
$\Gamma_S\left(\Omega_b({1\over2}^-)\to \Xi_b K\right) \neq 0$
&$\sim~0$&$\Omega_b(6350)$
\\ \cline{2-3} \cline{5-7}
&$\Omega_b({3\over2}^-)$&$6.353 \pm 0.129$&&--&$\sim~0$ & --
\\ \hline\hline
\end{tabular}
\label{tab:decaybottom}
\end{center}
\end{table*}

\begin{table*}[hbtp]
\begin{center}
\caption{Strong decay properties of the $P$-wave flavor-sextet charmed baryons, including the mixing between the $J^P=3/2^-$ members of the $[\mathbf{6}_F,1,1,\lambda]$ and $[\mathbf{6}_F,2,1,\lambda]$ multiplets with $\theta=(37\pm5)^\circ$. The broad $\rho$-mode $\Sigma_c$ and $\Xi_c^\prime$ states are omitted for compactness, and only non-negligible partial decay widths are quoted numerically. $\Gamma_S$ and $\Gamma_D$ denote the $S$- and $D$-wave partial widths, respectively. The small angles $\theta^\prime$ and $\theta^{\prime\prime}$ parameterize subleading mixing among other configurations with the same $J^P$, and decay channels opened by such mixing are marked by $\neq0$.}
\renewcommand{\arraystretch}{1.45}
\begin{tabular}{   c|c | c | c | c | c | c | c}
\hline\hline
  \multirow{2}{*}{HQET state}&\multirow{2}{*}{Mixing}&\multirow{2}{*}{Mixed state} & Mass & Splitting & Dominant decay modes & Width  & \multirow{2}{*}{Candidate}
\\  &&&   ({GeV}) & ({MeV}) & ({MeV})& ({MeV}) &
\\ \hline\hline
$[\Sigma_c({1\over2}^-),0,1,\lambda]$&\multirow{3}{*}{$\theta^\prime\approx 0^\circ$}&$[\Sigma_c({1\over2}^-),0,1,\lambda]$&$2.830^{+0.060}_{-0.040}$& -- &
$\begin{array}{l}
\Gamma_S\left(\Sigma_c({1\over2}^-)\to\Lambda_c \pi\right)=610^{+860}_{-410}
\end{array}$&$610^{+860}_{-410}$&--
\\ \cline{1-1}\cline{3-8}
$[\Sigma_c({1\over2}^-),1,1,\lambda]$&&$[\Sigma_c({1\over2}^-),1,1,\lambda]$&$2.832^{+0.172}_{-0.172}$&\multirow{4}{*}{$30^{+15}_{-20}$}&
$\begin{array}{l}
\Gamma_S\left(\Sigma_c({1\over2}^-)\to\Lambda_c \pi\right) \neq 0 \\
\Gamma_S\left(\Sigma_c({1\over2}^-)\to\Sigma_c\pi\right)=37^{+58}_{-27}\\
\end{array}$&$37^{+58}_{-27}$ & \multirow{6}{*}{$\Sigma_c(2800)$}
\\ \cline{1-4} \cline{6-7}
$[\Sigma_c({3\over2}^-),1,1,\lambda]$&\multirow{3}{*}{$\theta={37\pm5^\circ}$}&$|\Sigma_c({3\over2}^-)\rangle_L$&$2.862^{+0.167}_{-0.167}$&&
$\begin{array}{l}
\Gamma_D\left(\Sigma_c({3\over2}^-)\to\Lambda_c\pi\right)=20^{+65}_{-20}\\
\Gamma_D\left(\Sigma_c({3\over2}^-)\to\Sigma_c\pi\right)=7.7^{+37.8}_{-~7.7}\\
\Gamma_S\left(\Sigma_c({3\over2}^-)\to\Sigma_c^{*}\pi\right)=7.8^{+12.9}_{-~7.6}
\end{array}$&$35^{+116}_{-~35}$&\multirow{1}{*}{}
\\ \cline{1-1} \cline{3-7}
$[\Sigma_c({3\over2}^-),2,1,\lambda]$&&$|\Sigma_c({3\over2}^-)\rangle_H$&$2.879^{+0.228}_{-0.228}$&\multirow{4}{*}{$68^{+32}_{-35}$}&
$\begin{array}{l}
\Gamma_D\left(\Sigma_c({3\over2}^-)\to\Lambda_c\pi\right)=42^{+177}_{-~42}\\
\Gamma_S\left(\Sigma_c({3\over2}^-)\to\Sigma_c^{*}\pi\right)=4.4^{+6.3}_{-4.4}\\
\end{array}$&$46^{+183}_{-~46}$&\multirow{2}{*}{$\Sigma_c(2900)$}
\\ \cline{1-4} \cline{6-7}
$[\Sigma_c({5\over2}^-),2,1,\lambda]$&--&$[\Sigma_c({5\over2}^-),2,1,\lambda]$&$2.947^{+0.217}_{-0.217}$&&
$\begin{array}{l}
\Gamma_D\left(\Sigma_c({5\over2}^-)\to\Lambda_c\pi\right)=27^{+58}_{-20}\\
\Gamma_D\left(\Sigma_c({5\over2}^-)\to \Sigma_c\pi\right)=1.5^{+3.8}_{-1.3}\\
\Gamma_D\left(\Sigma_c({5\over2}^-)\to\Sigma_c^{*}\pi\right)=0.9^{+2.2}_{-0.8}
\end{array}$&$29^{+64}_{-22}$&\multirow{1}{*}{}
\\ \hline \hline
$[\Xi_c^\prime({1\over2}^-),0,1,\lambda]$&\multirow{4}{*}{$\theta^\prime \approx 0^\circ$}&$[\Xi_c^{\prime}({1\over2}^-),0,1,\lambda]$&$2.900^{+0.130}_{-0.120}$& -- &
$\begin{array}{l}
\Gamma_S\left(\Xi_c^{\prime}({1\over2}^-)\to\Lambda_c \bar K\right)=400^{+610}_{-270}\\
\Gamma_S\left(\Xi_c^{\prime}({1\over2}^-)\to \Xi_c \pi\right)=360^{+550}_{-250}
\end{array}$
&$760^{+820}_{-370}$&--
\\ \cline{1-1}\cline{3-8}
$[\Xi_c^\prime({1\over2}^-),1,1,\lambda]$&&$[\Xi_c^\prime({1\over2}^-),1,1,\lambda]$&$2.890^{+0.132}_{-0.132}$&\multirow{5}{*}{$35^{+15}_{-17}$}&
$\begin{array}{l}
\Gamma_S\left(\Xi_c^{\prime}({1\over2}^-)\to\Lambda_c \bar K\right) \neq 0 \\
\Gamma_S\left(\Xi_c^{\prime}({1\over2}^-)\to \Xi_c \pi\right)  \neq 0 \\
\Gamma_S\left(\Xi_c^{\prime}({1\over2}^-)\to \Xi_c^{\prime}\pi\right)=10^{+14}_{-~7}\\
\end{array}$&$10^{+14}_{-~7}$&$\Xi_c(2882)$
\\ \cline{1-4}\cline{6-8}
$[\Xi_c^{\prime}({3\over2}^-),1,1,\lambda]$&\multirow{4}{*}{$\theta={37\pm5^\circ}$}&$|\Xi_c^\prime({3\over2}^-)\rangle_L$&$2.925^{+0.128}_{-0.128}$&&
$\begin{array}{l}
\Gamma_D\left(\Xi_c^{\prime}({3\over2}^-)\to \Lambda_c \bar K\right)=1.8^{+2.7}_{-1.1}\\
\Gamma_D\left(\Xi_c^{\prime}({3\over2}^-)\to \Xi_c\pi\right)=4.0^{+8.6}_{-3.0}\\
\Gamma_D\left(\Xi_c^{\prime}({3\over2}^-)\to\Xi_c^{\prime}\pi\right)=1.6^{+2.0}_{-1.0}\\
\Gamma_S\left(\Xi_c^{\prime}({3\over2}^-)\to \Xi_c^{*}\pi\right)=2.0^{+2.9}_{-1.5}
\end{array}$&$9^{+16}_{-~7}$
& $\Xi_c(2923)$
\\ \cline{1-1}\cline{3-8}
$[\Xi_c^\prime({3\over2}^-),2,1,\lambda]$&&$|\Xi_c^\prime({3\over2}^-)\rangle_H$&$2.936^{+0.238}_{-0.238}$&\multirow{4}{*}{$64^{+30}_{-31}$}&
$\begin{array}{l}
\Gamma_D\left(\Xi_c^{\prime}({3\over2}^-)\to \Lambda_c \bar K\right)=4.3^{+6.3}_{-2.6}\\
\Gamma_D\left(\Xi_c^{\prime}({3\over2}^-)\to \Xi_c\pi\right)=9^{+18}_{-~6}\\
\Gamma_S\left(\Xi_c^{\prime}({3\over2}^-)\to \Xi_c^{*}\pi\right)= 1.2^{+1.8}_{-0.9}
\end{array}$&$15^{+26}_{-10}$&$\Xi_c(2939)$
\\ \cline{1-4} \cline{6-8}
$[\Xi_c^\prime({5\over2}^-),2,1,\lambda]$&--&$[\Xi_c^\prime({5\over2}^-),2,1,\lambda]$&$3.000^{+0.220}_{-0.220}$&&
$\begin{array}{l}
\Gamma_D\left(\Xi_c^{\prime}({5\over2}^-)\to \Lambda_c \bar K\right)=3.2^{+7.1}_{-2.4}\\
\Gamma_D\left(\Xi_c^{\prime}({5\over2}^-)\to \Xi_c \pi\right)=5.9^{+8.1}_{-3.4}\\
\Gamma_D\left(\Xi_c^{\prime}({5\over2}^-)\to \Xi_c^{*} \pi\right)=0.16^{+0.35}_{-0.16}
\end{array}$&$9^{+15}_{-~6}$
&$\Xi_c(2965)$
\\ \hline \hline
$[\Omega_c({1\over2}^-),0,1,\lambda]$&\multirow{2}{*}{$\theta^\prime \approx0^\circ$}&$[\Omega_c({1\over2}^-),0,1,\lambda]$&$3.030^{+0.180}_{-0.190}$& -- &$\begin{array}{l}
\Gamma_S\left(\Omega_c({1\over2}^-)\to\Xi_c \bar K\right)=980^{+1530}_{-~670}
\end{array}$&$980^{+1530}_{-~670}$&--
\\ \cline{1-1} \cline{3-8}
$[\Omega_c({1\over2}^-),1,1,\lambda]$&&$[\Omega_c({1\over2}^-),1,1,\lambda]$&$3.019^{+0.106}_{-0.106}$& \multirow{2}{*}{$31^{+14}_{-17}$} & $\Gamma_S\left(\Omega_c({1\over2}^-)\to\Xi_c \bar K\right) \neq 0$ &$\sim~0$&$\Omega_c(3000)$
\\ \cline{1-4} \cline{6-8}
$[\Omega_c({3\over2}^-),1,1,\lambda]$&\multirow{2}{*}{$\theta=37\pm5^\circ$}&$|\Omega_c({3\over2}^-)\rangle_L$&$3.050^{+0.100}_{-0.100}$&  &
$\begin{array}{l}
\Gamma_D\left(\Omega_c({3\over2}^-)\to \Xi_c \bar K\right)=1.3^{+2.6}_{-1.0}
\end{array}$&$1.3^{+2.6}_{-1.0}$&$\Omega_c(3050)$
\\ \cline{1-1} \cline{3-8}
$[\Omega_c({3\over2}^-),2,1,\lambda]$&&$|\Omega_c({3\over2}^-)\rangle_H$&$3.064^{+0.230}_{-0.230}$&\multirow{2}{*}{$55^{+27}_{-29}$}&$\begin{array}{l}
\Gamma_D\left(\Omega_c({3\over2}^-)\to \Xi_c \bar K\right)=3.7^{+7.3}_{-2.9}
\end{array}$&$3.7^{+7.3}_{-2.9}$&$\Omega_c(3066)$
\\ \cline{1-4} \cline{6-8}
$[\Omega_c({5\over2}^-),2,1,\lambda]$&--&$[\Omega_c({5\over2}^-),2,1,\lambda]$&$3.119^{+0.217}_{-0.217}$&&
$\begin{array}{l}
\Gamma_D\left(\Omega_c({5\over2}^-)\to \Xi_c \bar K\right)=3.3^{+6.4}_{-2.5}
\end{array}$&$3.3^{+6.4}_{-2.5}$&$\Omega_c(3090)$
\\ \hline
\multirow{1}{*}{$[\Omega_c({1\over2}^-),1,0,\rho]$}&\multirow{1}{*}{--}&\multirow{1}{*}{$[\Omega_c({1\over2}^-),1,0,\rho]$}&\multirow{1}{*}{$3.106^{+0.159}_{-0.159}$}& \multirow{3}{*}{$11^{+5}_{-5}$} &$\Gamma_S\left(\Omega_c({1/2}^-)\to\Xi_c^\prime \bar K\right) =260^{+540}_{-260}$ &\multirow{1}{*}{$260^{+540}_{-260}$}&\multirow{1}{*}{--}
\\ \cline{1-4} \cline{6-8}
\multirow{2}{*}{$[\Omega_c({3\over2}^-),1,0,\rho]$}&\multirow{2}{*}{$\theta^{\prime\prime} \approx0^\circ$}&\multirow{2}{*}{$[\Omega_c({3\over2}^-),1,0,\rho]$}&\multirow{2}{*}{$3.117^{+0.160}_{-0.160}$}&& $\Gamma_D\left(\Omega_c({3/2}^-)\to\Xi_c \bar K\right) \neq 0$ &\multirow{2}{*}{$0.3^{+0.5}_{-0.2}$}
&\multirow{2}{*}{$\Omega_c(3119)$}
\\ 
&&&&&$\Gamma_D\left(\Omega_c({3/2}^-)\to\Xi_c^\prime \bar K\right) =0.3^{+0.5}_{-0.2}$&
\\ \hline\hline
\end{tabular}
\label{tab:decaycharm}
\end{center}
\end{table*}

\begin{acknowledgments}
This project is supported by 
the National Natural Science Foundation of China under Grant No.~12075019, 
the Jiangsu Provincial Double-Innovation Program under Grant No.~JSSCRC2021488,
and 
the Fundamental Research Funds for the Central Universities.
\end{acknowledgments}

\bibliographystyle{elsarticle-num}
\bibliography{ref}

\end{document}